\documentclass[12pt]{article}

\usepackage{graphicx}    
\usepackage[dvips]{color}
\usepackage{epsfig,axodraw,epsf,rotate,graphicx}
\usepackage{cancel,pstricks,pst-node,pst-plot,psfrag,epic,eepic}

\def \gta {\mathrel{\vcenter {\hbox{$>$}\nointerlineskip\hbox{$\sim$}}}}

\def\bar{\overline}
\def\phi{\varphi}

\newcommand{\beq}{\begin{equation}}
\newcommand{\eeq}{\end{equation}}
\newcommand{\bea}{\begin{eqnarray}}
\newcommand{\eea}{\end{eqnarray}}
\newcommand{\be}{\begin{displaymath}}
\newcommand{\ee}{\end{displaymath}}
\newcommand{\bc}{\begin{center}}
\newcommand{\ec}{\end{center}}
\newcommand{\bt}{\begin{tabbing}}
\newcommand{\et}{\end{tabbing}}

\newcommand{\vs}{\vspace*{5mm}}
\newcommand{\vev}[1]{\left\langle #1\right\rangle}

\newcommand{\PCPC}{
\setlength{\unitlength}{1pt}
\begin{picture}(20,7)
\put(0,0){$P_{CP}$}
\setlength{\unitlength}{1cm}
\end{picture}}
\newcommand{\PCPV}{
\setlength{\unitlength}{1pt}
\begin{picture}(20,7)
\put(5,-2){\line(2,1){14}}
\put(0,0){$P_{CP}$}
\setlength{\unitlength}{1cm}
\end{picture}}

\newcommand{\reu}{{\nu_e\rightarrow\nu_\mu}}
\newcommand{\reub}{{\bar{\nu}_e\rightarrow\bar{\nu}_\mu}}
\newcommand{\ruu}{{\nu_\mu\rightarrow\nu_\mu}}
\newcommand{\ruub}{{\bar{\nu}_\mu\rightarrow\bar{\nu}_\mu}}
\newcommand{\rue}{{\nu_\mu\rightarrow\nu_e}}

\newcommand{\dm}[1]{{\Delta m^2_{#1}}}

\newcommand{\ie}{{\it i.e.}}

\newcommand{\eg}{{\it e.g.}}

\begin{document}


\begin{titlepage}

\renewcommand{\thefootnote}{\alph{footnote}}

\vspace*{-3.cm}
\begin{flushright}
TUM-HEP-474/02\\
\end{flushright}

\vspace*{2.5cm}

\begin{center}
{\Large\bf The Physics Potential of Future Long Baseline\\[4mm] 
Neutrino Oscillation Experiments} 
\end{center}

\vspace*{.8cm}
\vspace*{.3cm}
{\begin{center} {\large{\sc
                M.~Lindner\footnotetext{\makebox[1.cm]{Email:}
                lindner@ph.tum.de}\footnotetext{To appear in
                ``Neutrino Mass'', Springer Tracts in Modern Physics,  
                ed. by G. Altarelli and K. Winter.}
                }}
\end{center}}
\vspace*{0cm}
{\it
\begin{center}

Physik--Department, Technische Universit\"at M\"unchen,\\
James--Franck--Strasse, D--85748 Garching, Germany

\end{center}}

\vspace*{1.5cm}

{\Large \bf
\begin{center} Abstract \end{center}  }

We discuss in detail different future long baseline neutrino 
oscillation setups and we show the remarkable potential for 
very precise measurements of mass splittings and mixing angles.
Furthermore it will be possible to make precise tests of coherent
forward scattering and MSW effects, which allow to determine the 
sign of $\Delta m^2$. Finally strong limits or measurements of 
leptonic CP violation will be possible, which is very interesting 
since it is most likely connected to the baryon asymmtery 
of the universe.

\vspace*{.5cm}

\end{titlepage}

\newpage

\renewcommand{\thefootnote}{\arabic{footnote}}
\setcounter{footnote}{0}


\section{Introduction}
\label{sec:intro}

Since the existing evidence for atmospheric neutrino oscillations 
includes some sensitivity to the characteristic $L/E$ 
dependence of oscillations \cite{Toshito:2001dk}, there is 
little doubt that the observed flavour transitions are due to 
neutrino oscillations. Recently it has been established 
reliably that solar neutrinos undergo flavour transitions
\cite{Ahmad:2002jz,Ahmad:2002ka}, most likely due to oscillations.
This solves in any case the long standing solar neutrino problem, 
even though the characteristic $L/E$ dependence of oscillation 
is in this case not yet established. However, oscillation is 
under all alternatives by far the most plausible explanation 
and global oscillation fits to all available data clearly favour 
the so-called LMA solution for the mass splittings and mixings
\cite{Barger:2002iv,Bandyopadhyay:2002xj,Bahcall:2002hv,
deHolanda:2002pp}. The CHOOZ reactor experiment \cite{Apollonio:1999ae} 
provides moreover currently the most stringent upper bound for the 
sub-leading $U_{e3}$ element of the neutrino mixing matrix. 
The global pattern of neutrino oscillation parameters seems 
therefore quite well known and one may ask how precise future 
experiments will ultimately be able to measure mass splittings 
and mixings and what can be learned from such precise measurements.

The characteristic length scale $L$ of oscillations is given 
by $\dm{}L/E_\nu=\pi/2$ and the known atmospheric 
$\dm{31}$-value leads  thus to an oscillation length scale 
$L_{atm}$ as a function of energy. For  
$\dm{31} \simeq 3\cdot 10^{-3}$~eV and for neutrino energies 
of $E_\nu\simeq 10$~GeV one finds $L_{atm}\simeq {\cal O}(2000)$~km, 
\ie\ distances and energies which can be realized by sending 
neutrino beams from one point on the Earth to another. Such 
long baseline experiments (LBL) have the advantage that the 
source can in principle be controlled and understood very 
precisely. In contrast, natural neutrino sources like the sun 
or the atmosphere can not be controlled directly and they 
involve assumptions and indirect measurements. The precision of 
future solar and atmospheric oscillation experiments is thus at 
some level limited systematically by the source, which is in principle 
not the case for LBL experiments. The solar $\dm{21}$ is for the 
favoured LMA solution about two orders of magnitude smaller than 
the atmospheric $\dm{31}$, resulting for the same energies in 
oscillations at scales $L_{sol} \simeq (10-1000) \cdot L_{atm}$. 
The solar oscillations will thus not fully develop in such LBL
experiments on Earth, but sub-leading effects play nevertheless 
an important role in precision experiments. Another modification
comes from the fact that the neutrino beams of LBL experiments 
traverse the matter of the Earth. Coherent forward scattering 
in matter must therefore to be taken into account in precision 
experiments. This makes the analysis more involved, but as we will 
see, it offers also unique opportunities.

The existing K2K experiment \cite{Nakamura:tr} as well as MINOS 
\cite{Paolone:am} and CNGS \cite{Ereditato:an}, which are both 
under construction, are a promising first generation of LBL
experiments which will lead to improved oscillation parameters.
We will discuss in this article in some detail the remarkable 
potential of future LBL experiments and further details can be 
found in \cite{Huber:2002mx}. One important point is that the increased 
precision will allow to test in detail the three-flavouredness of 
oscillations. We will also see that it will be possible to limit or 
measure $\theta_{13}$ drastically better than today, that it is 
possible to study in detail MSW matter effects \cite{Wolfenstein:1978ue,
Wolfenstein:1979ni,Mikheev:1985gs,Mikheev:1986wj} and to extract in 
this way $sign(\dm{31})$, \ie\ the mass ordering of the neutrino states. 
For the now favoured LMA solution it will also be possible to measure 
leptonic CP violation. The precise neutrino masses, mixings and 
CP phases which can be obtained in this way are extremely valuable 
information about flavour physics, since unlike for quarks these 
parameters are not obscured by hadronic uncertainties. These parameters 
can then be evolved with the renormalization group \eg\ to the GUT 
scale\footnote{For examples see \cite{Antusch:2002hy,Antusch:2002rr}.}, 
where the rather precisely known parameters can be compared with mass 
models based on flavour symmetries or other models of neutrino and 
charged lepton masses. Leptonic CP violation is moreover related to 
leptogenesis \cite{FY:1986,Buchmuller:2001mw,Branco:2002kt}, the 
currently most plausible mechanism for the generation of the baryon 
asymmetry of the universe. LBL experiments offer therefore in a unique 
way precise knowledge on extremely interesting and valuable physics 
parameters.


\section{Beams and Detectors}
\label{sec:sod} 

Precise experiments in combination with an exact theoretical 
description are extremely valuable, since they allow precision 
determinations of the underlying quantities. Long baseline 
neutrino oscillation experiments are in principle of this type, 
since unlike for quarks there are no hadronic uncertainties
on theoretical side. Experimentally, LBL experiments have the 
advantage that both the source and the detector can be kept 
under precise conditions. This includes amongst others for the 
source a precise knowledge of the mean neutrino energy $E_\nu$, 
the neutrino flux and spectrum, as well as the flavour composition 
and contamination of the beam. Another important aspect is whether 
neutrino and anti-neutrino beam data can be obtained symmetrically 
such that systematical uncertainties cancel at least partly in an 
analysis. Precise measurements require also a sufficient luminosity
of the beam and a large enough detector such that sufficient 
statistics can be obtained. On the detector side, there are also 
a number of issues which have to be understood or determined very 
precisely, like the detection threshold function, energy calibration, 
resolution, particle identification capabilities (flavour, charge, 
event reconstruction, understanding backgrounds). Another source of 
uncertainty in the detection process is the knowledge of neutrino 
cross-sections, especially at low energies \cite{Paschos:2002mb}. 
The potential source and detector combinations of a future LBL 
experiment are furthermore constraint by the available technology. 
We will restrict ourself in the following to certain types of neutrino 
sources and detectors for LBL experiments. However, it is important 
to keep potential improvements of new source and detector developments 
in mind. An example is given by liquid argon detectors like ICARUS 
\cite{Arneodo:jt}. They are not included in this study, but they may 
become extremely valuable detectors or detector components in this context. 

\begin{figure}
\bc
\includegraphics[width=13cm]{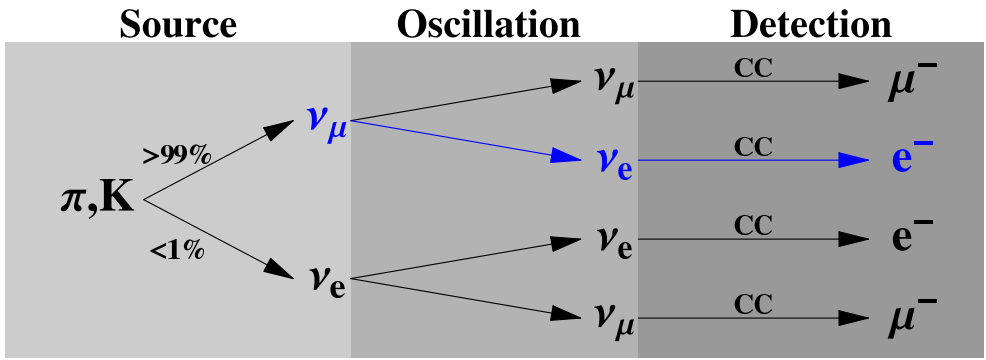} 
\caption{Schematic overview of neutrino production, oscillation and
detection via charged current interactions for conventional
beams and superbeams. The interesting channels are the $\ruu$ and 
$\rue$ disappearance and appearance channels. The $\nu_e$ beam 
contamination at the level of $<1\%$ limits the ability to 
determine the $\rue$ appearance oscillation, since it produces
also electrons without oscillation. The $\nu_\tau$ oscillation
channel is not shown here, but it would become very important
if tau lepton detection is feasible. This requires energies 
sufficient for tau production and detectors with suitable tau 
lepton detection capabilities.}
\label{fig:sbsignal}  
\ec
\end{figure}

The first type of considered sources are conventional neutrino and 
anti-neutrino beams. An intense proton beam is typically directed 
onto a massive target producing mostly pions and some K mesons, 
which are captured by an optical system of magnets in order to obtain 
a beam. The pions (and K mesons) decay in a decay pipe, yielding 
essentially a muon neutrino beam which can undergo oscillations as 
shown in fig.~\ref{fig:sbsignal}. Most interesting are the $\ruu$ 
disappearance channel and the $\rue$ appearance channels.
The neutrino beam is, however, contaminated by approximately
$0.5\%$ electron neutrinos, which also produce electron reactions 
in the disappearance channel, limiting thus the precision in the 
extraction of $\rue$ oscillation parameters. 
The energy spectrum of the muon beam can be controlled over 
a wide range: it depends on the incident proton energy, the 
optical system, and the precise direction of the beam axis 
compared to the direction of the detector. It is possible to 
produce broad band high energy beams, such as the CNGS 
beam~\cite{CNGS1,CNGS2}, or narrow band lower energy beams, 
such as in some configurations of the NuMI beam~\cite{MINOS}.
Reversing the electrical current in the lens system results 
in an anti-neutrino beam. The neutrino and anti-neutrino 
beams have significant differences such that errors do not 
cancel systematically in ratios or differences. The neutrino
and anti-neutrino beams must therefore more or less be considered 
as independent sources with different systematical errors.

So-called ``superbeams'' are based on the same beam dump techniques 
for producing neutrino beams, but at much larger luminosities 
\cite{CNGS1,CNGS2,MINOS,Nakamura:2000uu}. Superbeams are thus a
technological extrapolation of conventional beams, but use a 
proton beam intensity close to the mechanical stability limit of 
the target at a typical thermal power of $0.7\,\mathrm{MW}$ to 
$4\,\mathrm{MW}$. The much higher neutrino luminosity allows the 
use of the decay kinematics of pions to produce so--called 
``off--axis beams'', where the detector is located some degrees 
off the main beam axis. This reduces the neutrino flux and the 
average neutrino energy, but leads to a more mono-energetic beam 
and a significant suppression of the electron neutrino contamination. 
Several off--axis superbeams with energies of about $1\,\mathrm{GeV}$ 
to $2\, \mathrm{GeV}$ have been proposed in 
Japan~\cite{Itow:2001ee,Aoki:2002ks},
America~\cite{Para:2001cu}, and Europe~\cite{Gomez-Cadenas:2001eu,Dydak}.

The most sensitive neutrino oscillation channel for sub-leading 
oscillation parameters is the $\rue$ appearance transition. Therefore 
the detector should have excellent electron and muon charged current 
identification capabilities. In addition, an efficient rejection of 
neutral current events is required, because the neutral current 
interaction mode is flavor blind. With low statistics, the magnitude 
of the contamination itself limits the sensitivity to the $\rue$
transition severely, while the insufficient knowledge of its magnitude 
constrains the sensitivity for high statistics. A near detector 
allows a substantial reduction of the background 
uncertainties~\cite{Itow:2001ee,Szleper:2001nj} and plays a 
crucial role in controlling other systematical errors, such as the flux 
normalization, the spectral shape of the beam, and the neutrino cross 
section at low energies. At energies of about $1\,\mathrm{GeV}$, the 
dominant charge current interaction mode is quasi--elastic scattering, 
which suggests that water Cherenkov detectors are the optimal type 
of detector. At these energies, a baseline of about 
$300\,\mathrm{km}$ would be optimal to measure at the first
maximum of the oscillation. At about $2\,\mathrm{GeV}$, there is 
already a considerable contribution of inelastic scattering to the 
charged current interactions, which means that it would be useful to 
measure the energy of the hadronic part of the cross section. This 
favors low--Z hadron calorimeters, which also have a factor of ten 
better neutral current rejection capability compared to water 
Cherenkov detectors~\cite{Para:2001cu}. In this case, the optimum 
baseline is around $600\,\mathrm{km}$. The matter effects are expected
to be small for these experiments for two reasons. First of all, an
energy of about $1\,\mathrm{GeV}$ to $2\,\mathrm{GeV}$ is small 
compared to the MSW resonance energy of approximately $13\,\mathrm{GeV}$ 
in the upper mantle of the Earth. The second reason is that the baseline 
is too short to produce significant matter effects.

The second type of beam considered are so-called neutrino factories,
where muons are stored in the long straight sections of a storage ring.
The decaying muons produce muon and electron anti-neutrinos 
in equal numbers~\cite{Geer:1998iz}. The muons are produced by pion 
decays, where the pions are produced by the same technique as for 
superbeams. After being collected, they have to be cooled and 
reaccelerated very quickly. This has not yet been demonstrated and 
it is the major technological challenge for neutrino factories~\cite{FNAL}. 
The spectrum and flavor content of the beam are completely characterized 
by the muon decay and are therefore very precisely known~\cite{PDG}. 
The only adjustable parameter is the muon energy $E_\mu$, which is usually 
considered in the range from $20$ to $50\,\mathrm{GeV}$. In a neutrino 
factory it is also possible to produce and store anti-muons in order to 
obtain a CP conjugated beam. The symmetric operation of both beams leads 
to the cancellation or significant reduction of errors and systematical 
uncertainties. We will discuss in the following the neutrino beam, 
which implies always implicitly -- unless otherwise stated -- the 
CP conjugate channel.

\begin{figure}
\bc
\includegraphics[width=13cm]{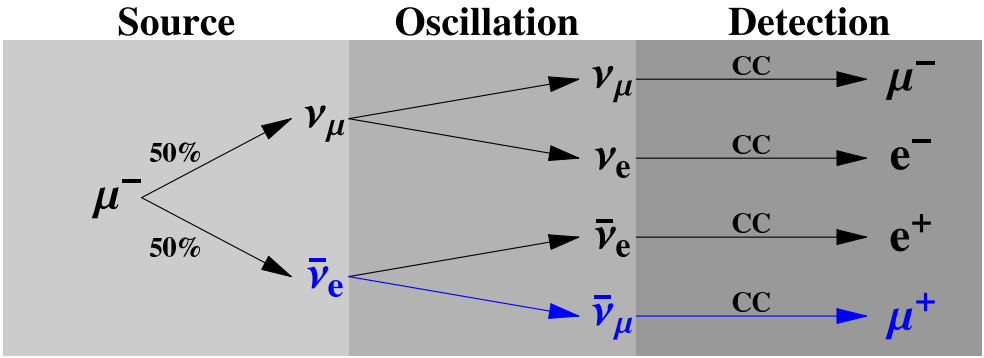} 
\caption{Neutrino production, oscillation and detection via 
charged current interactions for a neutrino factory for one
polarity. $\bar\nu_e$ and $\nu_\mu$ are produced in equal numbers
from $\mu$-decays and can undergo different oscillations. 
The $\ruu$ and $\reub$ channels are most interesting for 
detectors with $\mu$ identification. Note, however, that 
excellent charge identification capabilities are required to
separate ``wrong sign muons'' and ``right sign muons''.
The $\nu_\tau$ oscillation channel is not shown here, but
it would become important if detectors with tau identification
capabilities can be built.}
\label{fig:nufactsignal}  
\ec
\end{figure}

The decay of the muons and the relevant oscillation channels are shown 
in fig.~\ref{fig:nufactsignal}. 
Amongst all flavors and interaction types, 
muon charged current events are the easiest to detect. The appearance channel 
with the best sensitivity is thus the $\reub$ transition, which produces 
so called ``wrong sign muons''. Therefore, a detector must be able to very 
reliably identify the charge of a muon in order to distinguish wrong sign 
muons in the appearance channel from the higher rate of same sign muons in 
the disappearance channels. The dominant charge current interaction in the 
multi--GeV range is deep--inelastic scattering, making a good energy 
resolution for the hadronic energy deposition necessary. Magnetized iron 
calorimeters are thus the favored choice for neutrino factory detectors. 
In order to achieve the required muon charge separation, it is necessary 
to impose a minimum muon energy cut at 
approximately $4\,\mathrm{GeV}$~\cite{Blondel:2000gj,Albright:2000xi}. 
This leads to a significant loss of neutrino events in the range of about 
$4\,\mathrm{GeV}$ to $20\,\mathrm{GeV}$, which means that a high muon 
energy of $E_\mu = 50\,\mathrm{GeV}$ is desirable. The first oscillation 
maximum lies then at approximately $3\,000\,\mathrm{km}$. Matter effects 
are sizable at this baseline and energy and the limited knowledge of the 
Earth's matter density profile becomes an additional source of errors.

Finally so-called $\beta$-beams are an interesting type of beam.
The idea is to store radioactive isotopes in a storage ring similar 
to the muons in the neutrino factory, such that the $\beta$-decays 
of the radioactive elements lead to pure $\nu_e$ or $\bar\nu_e$ beams 
with $\gamma \simeq 100$ \cite{Zucchelli:2001gp}. 
The energy spectrum of the neutrinos in the beam is determined by the 
neutrino energies of the decay at rest, boosted by the $\gamma$ factor,
resulting typically in beam energies of a few hundred MeV at acceleration
energies of about $100$~GeV per nucleon. There are technological and 
environmental challenges and it is unclear if $\beta$-beams can become
an affordable and competitive neutrino source. We will not include 
$\beta$ beams in our quantitative discussion, but we will see that 
superbeams and neutrino factories have already an impressive potential, 
which could only be improved if $\beta$-beams were realized.


\section{The Oscillation Framework}
\label{sec:osc}

Most existing results on neutrino oscillations can so far be 
understood in an effective two neutrino framework with the well 
known oscillation probability for a baseline $L$ and neutrino 
energy $E_\nu$
\beq
P( \nu_{f_1} \rightarrow \nu_{f_2} ) =
\left|\vev{\nu_{f_1}(t)|\nu_{f_2}(t=0)}\right|^2 
=\sin^2 2\theta\cdot \sin^2\left(\frac{\Delta m^2 L}{4E_\nu}\right)~,
\label{eq:2osc}
\eeq
where $\theta$ is the mixing angle between the two flavour eigenstates 
$f_1$ and $f_2$ and where $\dm{}=m_2^2-m_1^2$ is the difference between 
the mass eigenvalues. Precision measurements at future LBL experiments 
involve a very precise knowledge of the sources, the detectors and the 
oscillation framework in matter. An effective two neutrino description 
is therefore definitively not adequate and matter effects must be included 
into the three neutrino oscillation framework. 

The generalization of the oscillation formulae in vacuum to the case 
of $N$ neutrinos leads to the probabilities for flavour transitions 
$\nu_{f_l} \rightarrow \nu_{f_m}$ given by
\beq
P( \nu_{f_l} \rightarrow \nu_{f_m} ) 
= \underbrace{\delta_{lm} - 4\sum_{i>j} \mathrm{Re} J_{ij}^{f_l
f_m}\sin^2\Delta_{ij}}_{{\mathbf\PCPC}}~ \underbrace{- 2\sum_{i>j} 
\mathrm{Im} J_{ij}^{f_l f_m}\sin 2\Delta_{ij}}_{{\mathbf\PCPV}}
\label{eq:Nosc}
\eeq
where the shorthands $J_{ij}^{f_l f_m} := U_{li}U_{lj}^*U^*_{mi}U_{mj}$
and $\Delta_{ij} := \frac{\Delta m^2_{ij} L}{4E}$ have been used. These
generalized vacuum transition probabilities depend on all combinations 
of quadratic mass differences $\Delta m^2_{ij}=m_i^2-m_j^2$ as well as 
on different products of elements of the leptonic mixing matrix $U$.

We will assume for the rest of this article a three neutrino framework 
which can easily be generalized to more neutrinos if necessary. We have
thus $1\leq i,j \leq 3$ and $U$ simplifies to the $3\times 3$ mixing 
matrix
\beq
U  =
\left(\begin{array}{ccc} 
c_{12}c_{13} & s_{12}c_{13} & s_{13}e^{-i\delta}\\
- s_{12}c_{23} - c_{12}s_{23}s_{13}e^{i\delta} & 
 c_{12}c_{23} - s_{12}s_{23}s_{13}e^{i\delta} & s_{23}c_{13}\\
s_{12}s_{23} - c_{12}c_{23}s_{13}e^{i\delta} 
& -c_{12}s_{23} - s_{12}c_{23}s_{13}e^{i\delta}
& c_{23}c_{13}\\ 
\end{array} \right)~,
\label{eq:MNS}
\eeq
which contains three leptonic mixing angles and one Dirac-like 
leptonic CP phase $\delta$. Note that the most general mixing matrix 
for three Majorana neutrinos contains two further Majorana-like 
CP phases. However, it can easily be seen that these two extra diagonal 
Majorana phases do not enter in the above oscillation formulae and 
therefore we can omit them safely. Three neutrino oscillations depend 
thus in general only on the three mixing angles and one CP-phase. 
Disappearance probabilities, \ie\ the transitions 
$\nu_{f_l} \rightarrow \nu_{f_l}$, do not even depend on this
CP-phase, since $J_{ij}^{f_l f_l}$ is only a function of the modulus 
of elements of $U$. Appearance probabilities, like 
$\nu_{e} \rightarrow \nu_{\mu}$ are therefore the place where
leptonic CP violation can be studied.

\begin{figure}[htb]
\begin{center}
\includegraphics[width=10cm]{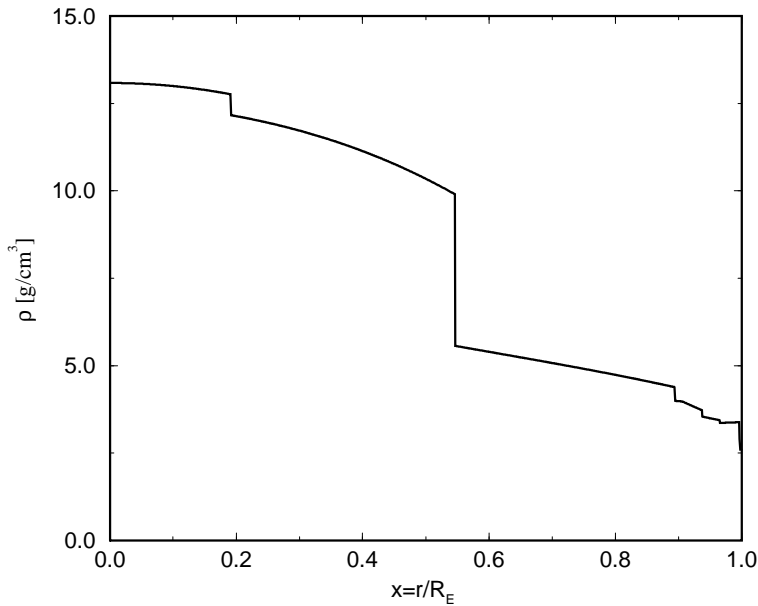}
\end{center}
\caption{The matter density profile of the Earth as a function of 
the radius $r$ \cite{Stacy:1977}. Such matter density profiles are 
obtained by combining different data coming from geology, material 
science, seismology and astronomy under the assumption of spherically 
symmetry.}
\label{fig:stacy}
\end{figure}

From eq.~(\ref{eq:Nosc}) the oscillation probabilities for neutrinos 
are $P(\nu_{f_l} \rightarrow \nu_{f_m} ) = \PCPC + \PCPV$ and 
for anti-neutrinos $P(\bar{\nu}_{f_l} \rightarrow \bar{\nu}_{f_m} ) 
= \PCPC - \PCPV$. Eq.~(\ref{eq:Nosc}) has thus a CP conserving part 
$\PCPC$, and a CP violating part $\PCPV$, and both terms depend on the
CP phase $\delta$. An extraction strategy for CP-violation seems 
thus given by looking at the CP asymmetries \cite{Dick:1999ed}
\beq
{a^{CP}:=\frac{
P(\nu_{f_l} \rightarrow \nu_{f_m} ) - P(\bar{\nu}_{f_l} \rightarrow \bar{\nu}_{f_m} )}{
P(\nu_{f_l} \rightarrow \nu_{f_m} ) + P(\bar{\nu}_{f_l} \rightarrow \bar{\nu}_{f_m} )} 
= \frac{^{\PCPV}}{_{\PCPC}}}~.
\label{eq:asym}
\eeq
Note, however, that the beams of a LBL experiment traverse the Earth
on a certain path and the presence of matter violates by itself CP, 
which modifies eq.~(\ref{eq:Nosc}) and which makes a measurement of 
leptonic CP violation more involved. The above general oscillation 
formulae in vacuum, eq.~(\ref{eq:Nosc}), lead to well known, but rather 
lengthy trigonometric expressions for the oscillation probabilities
in vacuum. These expressions become even longer and do not exist in 
closed form when arbitrary matter corrections are taken into account. 
However, the problem simplifies somewhat under the assumption of a 
spherically symmetric Earth matter distribution \cite{Stacy:1977} as shown 
in fig.~\ref{fig:stacy} as function of the Earth radius. 
There is a one to one correspondence between the baseline $L$ and the 
angle under which the beam must enter the Earth at the source. Some 
examples are given in table~\ref{tab:L} and obviously large $L$ 
corresponds to steep angles, resulting in technological and environmental 
challenges. 
\begin{table}
\bc
\begin{tabular}{|r|r|r|r|}
\hline
baseline $L$ in km & ~~2800~ & ~~7300~ & ~12750~ \\
\hline
angle in degrees & 13~ & 35~ & 90~ \\
\hline
\end{tabular}
\ec
\caption{Examples for oscillation lengths $L$ of long baseline 
experiments and the corresponding angle under which the beam 
enters and leaves the Earth near the source and near the detector.}
\label{tab:L}
\end{table}
Matter effects depend then in a first approximation only on the 
average matter density along the path between source and detector 
where fluctuations in the density profiles partially average out, 
but density errors must nevertheless be taken into account 
\cite{Geller:2001ix}. Matter effects in the oscillation formulae 
generally grow with distance $L$, while average density profile 
uncertainties tend to decrease with $L$, leading approximately to 
a constant error which we assume to be about $5\%$. Non-constant
matter profiles can in principle lead to very interesting oscillation effects
\cite{Akhmedov:kd,Petcov:1998su,Akhmedov:1998ui,Chizhov:1999az,%
Chizhov:1999he,Akhmedov:2000js}. Observing such effects would be
very interesting, but they do not affect the studied experiments and they
would just complicate the analysis of the oscillation parameter space.
Fig.~\ref{fig:stacy} shows at $r/R_E\simeq 0.5$ the pronounced 
density jump between the Earth mantle and the iron core and the 
average density approximation becomes much worse for beams which 
pass the core. Avoiding the iron core and the associated density 
jump corresponds to baselines of 
$L\leq {\cal O}(10\hspace*{2mm}000~\mathrm{km})$. 
Matter effects lead to an MSW resonance at a characteristic energy. 
It is interesting to note that this resonance energy lies for the 
atmospheric $\dm{31}$ and the given Earth matter density of about 
$3.8 \mathrm{g/cm}^3$ in the crust at $E_{\mathrm resonance}\simeq 
10-15~$GeV.

The addition of arbitrary matter effects would make the 
oscillation formulae in general rather complicated, but the 
problem becomes much simpler for the case of three neutrinos 
and approximately constant average matter density. The Hamiltonian 
describing three neutrino oscillation 
in matter can be written in flavour basis as
\beq
H =
\frac{1}{2E_\nu} U
\left(\begin{array}{ccc} 
m_1^2 & 0 & 0 \\ 
0 & m_2^2 & 0 \\ 
0 & 0 & m_3^2  
\end{array} 
\right) U^T
+
\frac{1}{2E_\nu} 
\left( \begin{array}{ccc} 
A+A' & 0 & 0\\ 
~0~ & ~A'~ & ~0~\\ 
~0~ & ~0~ & ~A'~ 
\end{array} 
\right) ~.
\label{eq:Hccnc}
\eeq
The first term describes oscillations in vacuum in flavour basis.
The quantities $A$ and $A'$ in the second term are given by the 
charged current and neutral current contributions to coherent 
forward scattering in matter. The charged current contribution 
is given by
\beq
A = \pm~ \frac{2\sqrt{2}G_F Y \rho E_\nu}{m_n}~,
\eeq
where $G_F$ is Fermi's constant, $Y$ is the number of electrons per 
nucleon, $m_n$ is the nucleon mass and $\rho$ is the matter density. 
A is positive for neutrinos in matter and anti-neutrinos in 
anti-matter, while it is negative for anti-neutrinos in matter 
and neutrinos in anti-matter. The flavour universal neutral current 
contributions $A'$ lead to an overall phase which does not enter the 
transition probabilities. The over-all neutrino mass scale $m_1^2$ can 
be written as a term proportional to the unit matrix and can similarly 
be removed, such that only $\dm{21}$ and $\dm{31}$ remain in the
first term of eq.~(\ref{eq:Hccnc}). Since $\dm{21} \ll \dm{31}$ we
may further approximately set $\dm{21}\simeq 0$ and we obtain thus 
the approximately equivalent Hamiltonian
\beq
H'  = 
\frac{1}{2E_\nu} U
\left(\begin{array}{ccc} 
~0~ & ~0~ & 0 \\ 
~0~ & ~0~ & 0 \\ 
~0~ & ~0~ & \dm{31}  
\end{array} 
\right) U^T
+
\frac{1}{2E_\nu} 
\left( \begin{array}{ccc} 
A & 0 & 0\\ 
0 & 0 & 0\\ 
0 & 0 & 0 
\end{array} 
\right) ~.
\label{eq:oscmatter}
\eeq
The mixing matrix $U$ can furthermore be written as a sequence of 
rotations $R_{ij}$ in the two dimensional sub-spaces $ij$, namely
\beq
U = R_{23}\cdot R_{13}\cdot R_{12}~,
\eeq
where the 12 and 23 rotations are real, \ie\ $R^T_{12}=R^{-1}_{12}$
and $R^T_{23}=R^{-1}_{23}$. This can be used to simplify the problem 
further, since $R_{12}$ commutes obviously with $diag(0,0,\dm{31})$ 
and the $\theta_{12}$ dependence disappears from eq.~(\ref{eq:oscmatter}). 
Next we observe that $diag(A,0,0)$ commutes with $R_{23}$, such that 
$R_{23}$ can be factored out from the complete Hamiltonian in 
eq.~(\ref{eq:oscmatter}), which can therefore be re-written as
\beq
H'  = R_{23}\left[
\frac{1}{2E_\nu} R_{13}
\left(\begin{array}{ccc} 
~0~ & ~0~ & 0 \\ 
~0~ & ~0~ & 0 \\ 
~0~ & ~0~ & \dm{31}  
\end{array} 
\right) R_{13}^T
+
\frac{1}{2E_\nu} 
\left( \begin{array}{ccc} 
A & 0 & 0\\ 
0 & 0 & 0\\ 
0 & 0 & 0 
\end{array} 
\right) 
\right] R_{23}^T~.
\label{eq:oscmatter2}
\eeq
Inside the square bracket of eq.~(\ref{eq:oscmatter2}) the original mass 
matrix is rotated by $R_{13}$. The matter dependent part is then added 
inside the square bracket, while $R_{23}$ was factored out. $H'$ is thus
diagonalized by $R_{23}R'_{13}$ and it becomes clear that we deal 
with a modification of the diagonalization of the 1-3 subspace. 
Matter effects lead therefore to an $A$-dependent parameter mapping 
in the 1-3 subspace which can be written as
\bea
\sin^2 2\theta_{13,m} &=& \frac{\sin^2 2\theta_{13}}{C_{\pm}^2} ~, \\
\dm{31,m} &=& \dm{31}C_{\pm} ~,\\
\dm{32,m} &=& \frac{\dm{31}~(C_{\pm}+1)+ A}{2}~, \\
\dm{21,m} &=& \frac{\dm{31}~(C_{\pm}-1)- A}{2}~,
\eea
where the index $m$ denotes effective quantities in matter and where
\beq
C^2_{\pm} = \left(\frac{A}{\dm{31}} 
-\cos 2\theta \right)^2 +\sin^2 2\theta~.
\label{eq:Cpm}
\eeq
Note that $A$ in $C_{\pm}$ can change its sign and the mappings for 
neutrinos and anti-neutrinos are therefore different, resulting
in different effective mixings and masses. This is an important 
effect, which will allow detailed tests of coherent forward 
scattering of neutrinos in matter at future LBL experiments. Note 
that oscillations in matter depend unlike vacuum oscillations 
via $C_{\pm}$ on the sign of $\dm{31}$. This is very interesting, 
since it opens the possibility to extract the $sign(\dm{31})$
via matter effects.

Another very interesting question is whether it will be possible 
to establish leptonic CP violation at future LBL experiments. 
Therefore we note that in the oscillation probabilities, 
eq.~(\ref{eq:Nosc}), all CP-violating effects are proportional 
to the following two quantities, namely
\bea
D & = & \sin\Delta_{21} \sin\Delta_{32} \sin\Delta_{31}~, 
\label{eq:D}\\
8 J_\mathrm{CP} & = & \cos\theta_{13} \sin(2\theta_{13}) 
    \sin(2\theta_{12}) \sin(2\theta_{23}) \sin\delta~.
\label{eq:JCP}
\eea
We can immediately see from eq.~(\ref{eq:D}) that CP violation is 
only possible if all three masses are different, \ie\ if none of the 
$\dm{ij}$ vanishes. For the considered LBL experiments we have 
$\Delta_{32} \simeq \Delta_{31} \simeq 1$ while $\sin \Delta_{21} 
\approx \Delta_{21} \ll 1$ and there is thus a suppression due to the 
small solar mass splitting. Furthermore we can see from eq.~(\ref{eq:JCP})
that $J_\mathrm{CP}$ is suppressed by the small value of $\theta_{13}$ 
and for the small mixing angle solution in addition by $\theta_{12}$. 
The largest CP violating effects occur thus when $\theta_{12}$
is large and for the largest possible solar $\dm{21}$, \ie\ the LMA 
solution, which interestingly happens to be the solution which is
clearly favoured by data. 

Putting everything together leads still to quite lengthy expressions 
for the oscillation probabilities in matter, where it is not easy to 
oversee all effects. It is therefore instructive to simplify the 
problem to a point, where an analytic understanding of all effects 
is possible, while quantitative statements should be obtained
with the help of numerical evaluations using the full expressions.
The key for further simplification is to expand the oscillation 
probabilities in small quantities. These expansion parameters are
$\alpha=\dm{21}/\dm{31} \simeq 10^{-2}$ and $\sin^2 2\theta_{13} \leq 0.1$.
The matter effects can be parametrized by the dimensionless quantity
$\hat A=A/\dm{31}=2VE/\dm{31}$, where $V=\sqrt{2}G_F n_e$.
Using $\Delta \equiv \Delta_{31}$, the leading terms in this expansion 
are, for example, for $P(\nu_\mu \rightarrow \nu_\mu)$ and 
$P(\nu_e \rightarrow \nu_\mu)$ \cite{Cervera:2000kp,Freund:2001pn,Freund:2001ui}
\begin{eqnarray}
& & \hspace*{-8mm}P(\nu_\mu \rightarrow \nu_\mu)  \approx \nonumber \\
& & 
1 - \cos^2 \theta_{13} \sin^2 2\theta_{23} \sin^2 {\Delta}
+ 2~{\alpha}~  \cos^2 \theta_{13} \cos^2 \theta_{12} \sin^2
2\theta_{23} {\Delta} \cos{\Delta},
\label{eq:Pdis}\\
\nonumber \\
& & \hspace*{-8mm}P(\nu_e \rightarrow \nu_\mu) \approx 
{\sin^2 2\theta_{13}}\sin^2 \theta_{23}~
 ~\frac{\sin^2({(1\!\!-\!\hat{A})}{\Delta})}{{(1\!\!-\!\hat{A})^2}} \nonumber\\
& & \pm~
{\sin\delta}\cdot {\sin 2\theta_{13}}~{\alpha}~ \sin 2\theta_{12} \cos\theta_{13}
   \sin 2\theta_{23}\sin({\Delta})\frac{
\sin({\hat{A}}{\Delta})\sin({(1\!\!-\!\hat{A})}{\Delta})}{{\hat{A}(1\!\!-\!\hat{A})}}
\nonumber \\
& &+ \cos\delta\cdot {\sin 2\theta_{13}}~ {\alpha}~  \sin 2\theta_{12} \cos\theta_{13}  
\sin 2\theta_{23}\cos({\Delta})\frac{\sin({
\hat{A}}{\Delta})\sin({(1\!\!-\!\hat{A})}{\Delta})}{{\hat{A}(1\!\!-\!\hat{A})}}
\nonumber \\
& &+~  {\alpha^2}~ \sin^2 2\theta_{12} \cos^2 \theta_{23} 
\frac{\sin^2({\hat{A}}{\Delta})}{{\hat{A}^2}}~,
\label{eq:Pap}
 \end{eqnarray}
where in eq.~(\ref{eq:Pap}) ``$+$'' stands for neutrinos and ``$-$'' 
for anti-neutrinos.
The most important feature of eq.~(\ref{eq:Pap}) is that all interesting 
effects in the $\reu$ transition depend crucially on $\theta_{13}$. The 
size of $\sin^2 2\theta_{13}$ determines thus if the total transition rate, 
matter effects, effects due to the sign of $\dm{31}$ and CP violating
effects are measurable. One of the most important questions for future 
LBL experiments is therefore how far experiments can push the 
$\theta_{13}$ limit below the current CHOOZ bound of approximately  
$\sin^2 2\theta_{13}<0.1$.


Before we discuss further important features of eqs.~(\ref{eq:Pdis}) 
and (\ref{eq:Pap}) in more detail we would like to comment once more
on the underlying assumptions and the reliability of these equations.
First eqs.~(\ref{eq:Pdis}) and (\ref{eq:Pap}) are an expansion in 
terms of the small quantities $\alpha$ and $\sin 2\theta_{13}$.
Higher order terms are suppressed at least by another power of 
one of these small parameters and these corrections are thus typically
at the percent level. The matter corrections in eqs.~(\ref{eq:Pdis}) 
and (\ref{eq:Pap}) are derived for constant average matter density.
Numerical test have shown that this approximation works quite well
as long as the matter profile is reasonably smooth. A number of
very interesting effects existing in general non-constant matter 
distributions are therefore only small theoretical uncertainties. 
An example is given by asymmetric matter profiles, which lead to 
interesting T-violating effects \cite{Akhmedov:2001kd}, but this 
does not play a role here since the Earth is sufficiently symmetric. 

Note that all results which will be shown later are based on 
numerical simulations of the full problem in matter. These results 
do therefore not depend on any approximation. Eqs.~(\ref{eq:Pdis}) 
and (\ref{eq:Pap}) will only be used to understand the problem 
analytically, which is extremely helpful in order to oversee the 
six (or more) dimensional parameter space. The full numerical 
analysis and eqs.~(\ref{eq:Pdis}) and (\ref{eq:Pap}) depend,
however, on the assumption of a standard three neutrino scenario. 
It is thus assumed that the LSND signal \cite{Church:2002tc}
will not be confirmed by the MiniBooNE experiment \cite{Hawker:pt}.
Neutrinos could in principle decay, which would make the analysis 
much more involved. It is assumed in this article that neutrinos 
are stable, and a combined treatment of oscillation and decay 
\cite{Lindner:2001fx} would be much more involved. Neutrinos might 
further have unusual properties and might, for example, violate CPT. 
In that case neutrinos and anti-neutrinos could have different 
properties and LBL experiments can give very interesting limits 
on this possibility \cite{Bilenky:2001ka}, but we will assume in 
this study that CPT is preserved.


\section{Correlations and Degeneracies}
\label{sec:CD}

Eqs.~(\ref{eq:Pdis}) and (\ref{eq:Pap}) exhibit certain parameter 
correlations and degeneracies, which play an important role in the 
analysis of LBL experiments, and which would be hard to understand
in a purely numerical analysis of the high dimensional parameter 
space. The most important properties are:

\begin{itemize}
\item
First we observe that eqs.~(\ref{eq:Pdis}) and (\ref{eq:Pap})
depend only on the product $\alpha\cdot \sin 2\theta_{12}$ or 
equivalently $\dm{21}\cdot \sin 2\theta_{12}$. This are the 
parameters related to solar oscillations which will be taken
as external input. The fact that only the product enters, implies
that it may be better determined than the product of the 
measurements of $\dm{21}$ and $\sin 2\theta_{12}$. 
\item
Next we observe in eq.~(\ref{eq:Pap}) that the second and third 
term contain both a factor $\sin(\hat A\Delta)$, while the last 
term contains a factor $\sin^2(\hat A\Delta)$. Since 
$\hat A\Delta= 2VL$, we find that these factors depend only on 
$L$, resulting in a ``magic baseline'' when $2VL_{magic}=\pi/4V$, 
where $\sin(\hat A\Delta)$ vanishes. At this magic baseline only 
the first term in eq.~(\ref{eq:Pap}) survives and 
$P(\nu_e \rightarrow \nu_\mu)$ does no longer depend on $\delta$, 
$\alpha$ and $\sin 2\theta_{12}$. This is in principle very 
important, since it implies that $\sin^2 2\theta_{13}$ can be 
determined at the magic baseline from the first term of 
eq.~(\ref{eq:Pap}) whatever the values and errors of $\delta$, 
$\alpha$ and $\sin 2\theta_{12}$ are. For the matter density of 
the Earth we find 
\beq
L_{magic}= \pi/4V \simeq 8100~\mathrm {\rm km}~,
\label{eq:Lmagic}
\eeq
which is an amazing number, since the value of $V$ could be such 
that $L_{magic}$ is very different from the scales under
discussion.
\item
Next we observe that only the second and third term of 
eq.~(\ref{eq:Pap}) depend on the CP phase $\delta$, and 
both terms contain a factor $\sin 2\theta_{13}\cdot\alpha$, 
while the first and fourth term of eq.~(\ref{eq:Pap})
do not depend on the CP phase $\delta$ and contain factors of  
$\sin^2 2\theta_{13}$ and $\alpha^2$, respectively.
The extraction of CP violation is thus always suppressed by 
the product $\sin 2\theta_{13}\cdot\alpha$ and the CP violating 
terms are furthermore obscured by large CP independent terms if 
either $\sin^2 2\theta_{13} \ll \alpha^2$ or
$\sin^2 2\theta_{13} \gg \alpha^2$. The determination of the 
CP phase $\delta$ is thus best possible if 
$\sin^2 2\theta_{13} \simeq 4\theta^2_{13} \simeq \alpha^2$.
\item
Another observation is that the last term in eq.~(\ref{eq:Pap}),
which is proportional to $\alpha^2=(\dm{21})^2/(\dm{31})^2$,
dominates in the limit of tiny $\sin^2 2\theta_{13}$. The error 
of $\dm{21}$ limits therefore for small $\sin^2 2\theta_{13}$ the 
parameter extraction.
\item
Eqs.~(\ref{eq:Pdis}) and (\ref{eq:Pap}) have a structure which 
suggests that transformations exist, which leave these equations 
invariant. We expect therefore degeneracies, \ie\ for given 
$L/E$ parameter sets with identical oscillation probabilities. 
An example of such an invariance is given by a simultaneous replacement
of neutrinos by anti-neutrinos and $\dm{31}\rightarrow -\dm{31}$. 
This is equivalent to changing the sign of the second term of 
eq.~(\ref{eq:Pap}) and replacing $\alpha \rightarrow -\alpha$ and
$\Delta \rightarrow -\Delta$, while $\hat A \rightarrow \hat A$.
It is easy to see that eqs.~(\ref{eq:Pdis}) and (\ref{eq:Pap}) are
unchanged, but this constitutes no degeneracy, since we can 
distinguish neutrinos and anti-neutrinos experimentally. 
\item
The first real degeneracy \cite{Barger:2001yr} can be seen in the 
disappearance probability eq.~(\ref{eq:Pdis}), which is invariant 
under the replacement $\theta_{23}\rightarrow \pi/2 - \theta_{23}$.
Note that the second and third term in eq.~(\ref{eq:Pap}) are not 
invariant under this transformation, but this change in the sub-leading 
appearance probability can approximately be compensated by small 
parameter shifts. However, the degeneracy can in principle be lifted
with precision measurements in the disappearance channels.
\item
The second degeneracy can be found in the appearance probability 
eq.~(\ref{eq:Pap}) in the ($\delta-\theta_{13}$)-plane 
\cite{Burguet-Castell:2001ez}. In terms of $\theta_{13}$ (which is 
small) and $\delta$ the four terms of eq.~(\ref{eq:Pap}) have the 
structure  
\beq
P(\nu_e \rightarrow \nu_\mu) \approx 
\theta^2_{13}\cdot F_1 +
\theta_{13}\cdot (\pm~\sin\delta F_2 + \cos\delta F_3) +
F_4~,
\label{eq:degth13}
\eeq
where the quantities $F_i$, $i=1,..,4$ contain all the other parameters. 
The requirement $P(\nu_e \rightarrow \nu_\mu)=const.$ leads for both 
neutrinos and anti-neutrinos to parameter manifolds of degenerate
or correlated solutions. Having both neutrino and anti-neutrino beams, 
the two channels can be used independently, which is equivalent to 
considering simultaneously eq.~(\ref{eq:degth13}) for $F_2\equiv 0$ 
and $F_3\equiv 0$. The requirement that these probabilities are now 
independently constant, \ie\ $P(\nu_e \rightarrow \nu_\mu)=const.$ 
for $F_2\equiv 0$ and $F_3\equiv 0$, leads to more constraint manifolds 
in the ($\delta-\theta_{13}$)-plane, but some degeneracies still survive.
\item
The third degeneracy \cite{Minakata:2001qm} is given by the fact 
that a change in sign of $\dm{31}$ can essentially be compensated 
by an offset in $\delta$. Therefore we note again that the transformation 
$\dm{31} \rightarrow -\dm{31}$ leads to $\alpha \rightarrow -\alpha$,
$\Delta \rightarrow -\Delta$ and $\hat A \rightarrow -\hat A$. 
All terms of the disappearance probability, eq.~(\ref{eq:Pdis}), are 
invariant under this transformation. The first and fourth term in 
the appearance probability eq.~(\ref{eq:Pdis}), which do not depend
on the CP phase $\delta$, are also invariant. The second and third 
term of eq.~(\ref{eq:Pdis}) depend on the CP phase and change by the 
transformation $\dm{31} \rightarrow -\dm{31}$. The fact that these 
changes can be compensated by an offset in the CP phase $\delta$ 
is the third degeneracy.
\item
Altogether there exists thus an eight-fold degeneracy 
\cite{Barger:2001yr}, as long as only the $\ruu$, $\ruub$,
$\reu$ and $\reub$ channels and one fixed $L/E$ are considered. 
However, the structure of eqs.~(\ref{eq:Pdis}) and (\ref{eq:Pap}) 
makes clear that the degeneracies can be broken by using in a 
suitable way information from different $L/E$ values. This can be 
achieved in total event rates by changing $L$ or $E$ 
\cite{Burguet-Castell:2002qx,Barger:2002rr}, but it can in principle 
also be done by using information in the event rate spectrum of a
single baseline $L$, which requires detectors with very good energy 
resolution \cite{Freund:2001ui}. Another strategy to break the 
degeneracies is to include further oscillation channels in the 
analysis (``silver channels'') \cite{Donini:2002rm,Burguet-Castell:2002qx}.
\end{itemize}
The discussion of this section shows the strength of the analytic 
approximations, which allow to understand the complicated parameter 
interdependence. It also helps to optimally plan experimental 
setups and to find strategies to resolve the degeneracies.


\section{Event Rates}
\label{sec:evrates} 

The experimentally detected event rates must be compared with 
the theoretical expressions, which depend only indirectly on 
the above oscillation probabilities. Every event can be classified 
by the information on the flavor of the detected neutrino and the 
type of interaction. The particles detected in an experiment are 
produced by neutral current (NC), inelastic charged current (CC) 
or quasi--elastic charged current (QE) interactions. The contribution 
to each mode depends on a number of factors, like detector type, the 
neutrino energy and flavour. In order to calculate realistic event
rates we compute first for each neutrino flavor and energy bin
the number of events for each type of interaction in the fiducial 
mass of an ideal detector. Next the deficiencies of a real detector
are included, like limited event reconstruction capabilities. 
The combined description leads to the differential event rate 
spectrum for each flavor and interaction mode as it would be seen 
by a detector which is able to separate all these channels.
Finally different channels must be combined, since they can not
be observed separately. This can be due to physics, \eg, due to
the flavor--blindness of NC interactions, or it can be a consequence
of detector properties, \eg, due to charge misidentification.
The differential event rates can thus be written for each channel 
of the interaction type IT as
\begin{eqnarray}
\label{eq:master}
\frac{dn_{f}^{\mathrm{IT}}}{dE'}=&&N\,\sum_{i}\int \int dE\,d\hat{E}\quad
\underbrace{\Phi_{i} (E)}_{\mathrm{Production}} \times \nonumber\\
&&\underbrace{\frac{1}{L^2} P_{(i\rightarrow f)}
(E,L,\rho;\theta_{23},\theta_{12},\theta_{13},
\Delta m^2_{31},\Delta m^2_{21},\delta)}_{\mathrm{Propagation}}~\times \nonumber \\ 
&&\underbrace{\sigma^{\mathrm{IT}}_f(E)
k_f^{\mathrm{IT}}(E-\hat{E})}_{\mathrm{Interaction}} ~~\times~~ 
\underbrace{ T_f(\hat{E}) V_f(\hat{E}-E')}_{\mathrm{Detection}}~,
\label{eq:EVrates} 
\end{eqnarray}
where $f$ and $i$ stand for the final and initial neutrino flavor, 
respectively. $E$ is the incident neutrino energy, $\Phi_{i} (E)$ is 
the flux for the initial flavor $i$ from the source, $L$ is the 
baseline, $N$ is a normalization factor, and $\rho$ is the Earth matter 
density. The interaction term is composed of two factors, which
are the total cross section $\sigma^{\mathrm{IT}}_f(E)$ for the flavor 
$f$ and the interaction type IT, and the energy distribution of the 
secondary particle $k_f^{\mathrm{IT}}(E-\hat{E})$, where $\hat{E}$ is
the energy of the secondary particle. The detector threshold is 
parametrized by the function $T_f(\hat{E})$, describing limited 
resolution or cuts in the analysis. The energy resolution of the
detector is parametrized by the function $V_f(\hat{E}-E')$ for 
the secondary particle, where $E'$ is the reconstructed neutrino energy.

The numerical calculation of the double integral of eq.~(\ref{eq:EVrates}) 
for all possible parameter combinations requires enormous computing power.
We use therefore an approximation where we evaluate the integral over
$\hat{E}$, where the only terms containing $\hat{E}$ are
$k_f^{\mathrm{IT}}(E-\hat{E})$,  $ T_f(\hat{E})$, and $ V_f(\hat{E}-E')$.
We define
\begin{eqnarray} R_f^{\mathrm{IT}}(E,E')\,\epsilon_f^{\mathrm{IT}}(E') \equiv
\int d\hat{E} \quad T_f(\hat{E})\,k_f^{\mathrm{IT}}(E-\hat{E})
\,V_f(\hat{E}-E')~,
\label{eq:EVapprox1} 
\end{eqnarray}
and approximate $R_f^{\mathrm{IT}}$ by the analytical expression
\beq
R_f^{\mathrm{IT}}(E,E')=
\frac{1}{\sigma\sqrt{2\pi}}\exp{\frac{(E-E')^2}{2\sigma^2}}~.
\label{eq:EVapprox2}
\eeq
For QE interactions in JHF beam we will use \cite{Itow:2001ee} 
$\sigma= 85$~MeV and for the neutrino factory and NuMI beams 
we use \cite{Ables:1995wq,Para:priv} $\sigma = 0.15\cdot E$. The 
values for the effective relative energy resolution $\delta E$ 
and the effective efficiency $\epsilon_f^{\mathrm{IT}}$ can be 
found in the literature, \ie\ for the neutrino factory in 
\cite{Blondel:2000gj,Albright:2000xi,Cervera:2000kp,Agafonova:2000xm,
Cervera:2000vy} and for the superbeam setups in \cite{Itow:2001ee,
SK1,SK2,SK3,SK4}. The threshold for muon detection is for neutrino 
factories an important parameter and we use essentially an 
interpolation between a more optimistic and a conservative 
attitude \cite{Albright:2000xi,Cervera:2000kp}. For further 
details see \cite{Huber:2002mx}.

In order to include backgrounds, the channels are grouped in an experiment 
specific way into pairs of signal and background. The considered backgrounds 
are NC--events which are misidentified as CC--events and CC--events identified 
with the wrong flavor or charge. For superbeams we include furthermore the 
background of CC--events coming from an intrinsic contamination of the beam. 

Finally we combine in the analysis all available signal channels and 
perform a global fit to extract the physics parameters in an optimal way.
The relevant channels are for a neutrino factory for each polarity of the 
beam the $\nu_\mu$--CC channel (disappearance) and $\bar{\nu}_\mu$--CC channel 
(appearance) event rate spectra. The backgrounds for these signals are 
NC events for all flavours and misidentified $\nu_\mu$--CC events.
For superbeam experiments the signal is for each polarity of the beam 
given by the $\nu_\mu$--QE channel (disappearance) and ${\nu}_e$--CC 
channel (appearance). The backgrounds are here NC events for all flavors, 
misidentified $\nu_\mu$--CC events, and, for the  ${\nu}_e$--CC channel, 
the $\nu_e$--CC beam contamination.


\section{The Considered LBL Setups}
\label{sec:LBLpot}

The discussed sources and detectors allow different LBL experiments 
and it is interesting to compare their physics potential on an 
equal and as realistic as possible footing. Studies at
the level of probabilities are not sufficient and the true potential 
must be evaluated at the level of event rates as described in 
section~\ref{sec:evrates}, with realistic assumptions about the 
beams, detectors and backgrounds. We present now the results of such 
an analysis which is essentially based on reference \cite{Huber:2002mx}, 
where we calculate the oscillation probabilities with 
the exact three neutrino oscillation formulae in matter numerically, 
\ie\ we use the approximations for the probabilities in eqs.~(\ref{eq:Pdis}) 
and (\ref{eq:Pap}) only for a qualitative understanding. All results 
shown below are therefore not affected by approximations which were 
made in the derivation of the approximate analytic oscillation formulae 
eqs.~(\ref{eq:Pdis}) and (\ref{eq:Pap}).
Sensitivities etc. are defined by the ability to re-extract the 
physics parameters from a simulation of event rates. Therefore
event rate distributions are generated for any possible parameter 
set. Subsequently a combined fit to these event rate distributions 
is performed simultaneously for the appearance and disappearance 
channels for both polarities. This procedure uses all the available 
information in an optimal way. It includes spectral distributions when 
present, and it reduces to a fit of total rates when the total event 
rates are small. Adequate statistical methods as described in 
\cite{Huber:2002mx} are used in order to deal with event distributions 
which have in some regions small event rates per bin. Systematical 
uncertainties are parametrized and external input from geophysics is used 
in form of the detailed matter profile and its errors, which are included 
in the analysis \cite{Huber:2002mx}. The ability to re-extract in a 
simulation of the full experiment the input parameters which were 
used to generate event rate distributions is used define sensitivities 
and precision.

An important aspect of such an analysis is the inclusion of external 
information. The discussed LBL experiments could in principle measure 
the solar $\dm{21}$ and mixing angle $\theta_{12}$. However, the precision 
which can be obtained can not compete with the expected measurement of 
KamLand \cite{Barger:2000hy}. We 
include therefore as external input the assumption that KamLand measures the 
solar parameters in the center of the LMA region with typical errors. 
Otherwise all unknown parameters (like the CP phase) will not be constrained
and are therefore left free, with all parameter degeneracies and error correlations
taken into account. All nuisance parameters are integrated out and a projection 
on the parameter of interest is performed. Altogether we are dealing with 
six free parameters.

\begin{figure}[tb]
\bc
\includegraphics[width=7.3cm]{./jhfflux.eps} 
\caption{The flux of the $2^o$ off-axis JHF beam as a function of 
energy. The mean energy is $0.51\,\mathrm{GeV}$ and the peak intensity is 
$1.7\cdot10^7\,\mathrm{GeV}^{-1}\,\mathrm{cm}^{-2}\,\mathrm{yr}^{-1}$ 
at $0.78\,\mathrm{GeV}$. The $\nu_e/\nu_\mu$-ratio is at the peak $0.2\%$.}
\label{fig:JHFbeam}
\ec
\end{figure}

\begin{figure}[htb]
\bc
\includegraphics[width=7.3cm]{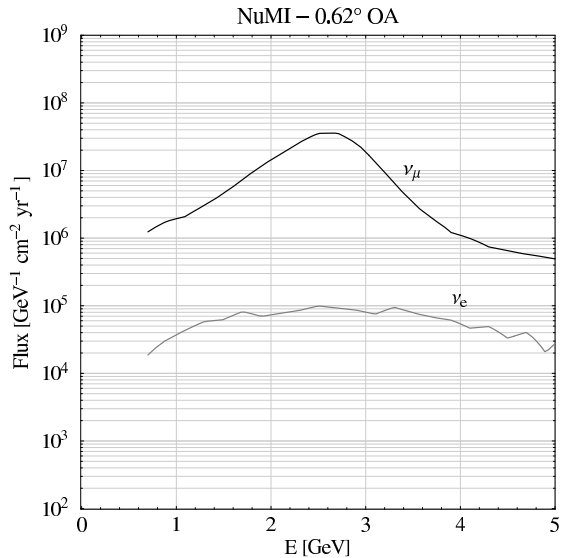} 
\caption{The flux of the proposed NuMI off-axis beam with a mean energy of 
$2.78\,\mathrm{GeV}$, a peak intensity of 
$3.6\cdot10^7\,\mathrm{GeV}^{-1}\,\mathrm{cm}^{-2}\,\mathrm{yr}^{-1}$ at
$2.18\,\mathrm{GeV}$. The $\nu_e/\nu_\mu$-ratio is at the peak $0.2\%$.}
\label{fig:NUMIbeam}
\ec
\end{figure}

\begin{figure}[htb]
\bc
\includegraphics[width=7.3cm]{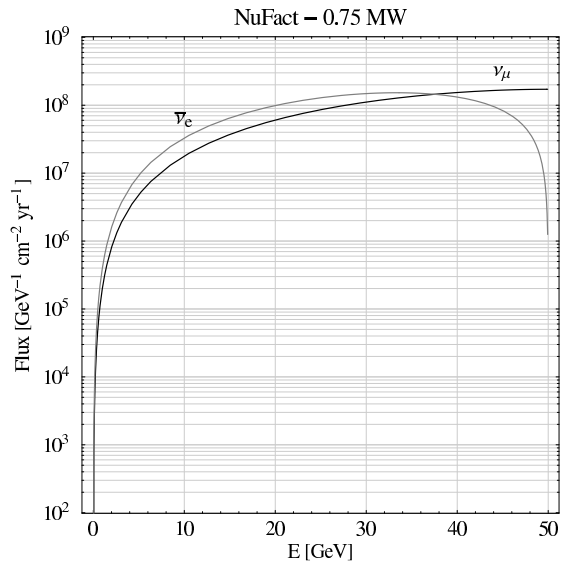} 
\caption{The flux of a neutrino factory with $E_\mu=50~\mathrm{GeV}$.
The mean neutrino energy is about $30\,\mathrm{GeV}$, the peak intensity
is $1.5\cdot10^8\,\mathrm{GeV}^{-1}\,\mathrm{cm}^{-2}\,\mathrm{yr}^{-1}$
at $33.33\,\mathrm{GeV}$. The $\nu_\mu/\nu_e$-ratio is at the peak $83\%$.}
\label{fig:NFbeam}
\ec
\end{figure}

The beam characteristics of the three considered sources are shown in
figures~\ref{fig:JHFbeam} (JHF), \ref{fig:NUMIbeam} (NuMI off-axis) 
and \ref{fig:NFbeam} (neutrino factory). We also include uncertainties
of these beam parameters, \ie\ for the first two conventional beams 
uncertainties in the $\nu_e$--background and for all beams flux 
uncertainties \cite{Itow:2001ee,Para:2001cu,Geer:1998iz}.
As detectors we consider water Cherenkov detectors, low-Z 
calorimeters and magnetized iron detectors with parameters as given 
in table~\ref{tab:detectors}. For magnetized iron calorimeters it is
important to include realistic threshold effects. We use a linear 
rise of the efficiency between $4\,\mathrm{GeV}$ and $20\,\mathrm{GeV}$
and we study the sensitivity to the threshold position. 
We do not include liquid Argon TPCs in our analysis, but they would 
certainly be an important detector if this technology will work.
\begin{table}
\bc
\begin{tabular}{|l|c|c|c|}
\hline
 &  {water Cherenkov}  & {low-Z}  & {magnetized iron}\\
 &  {= SK (HK)} &  {calorimeter} & {calorimeter}\\
\hline
fiducial mass & $22.5\,\mathrm{kt}$ ($1\,000\,\mathrm{kt}$) & $20\,\mathrm{kt}$ & $10\,\mathrm{kt}$ ($50\,\mathrm{kt}$)\\
energy range & $0.4-1.2\,\mathrm{GeV}$ & $1-5\,\mathrm{GeV}$ & $4-50\,\mathrm{GeV}$\\
energy resolution & $85$~MeV & $0.15\cdot E$ & $0.15\cdot E$\\
signal efficiency & $0.5$ & $0.5$ & $0.45$\\
NC rejection & $0.01$ & $0.001$ & $<10^{-5}$\\
CID & -- & -- & $< 10^{-5}$\\
background uncertainty  &  $5\%$  &  $5\%$  &  $5\%$ \\
\hline
\end{tabular}
\ec
\caption{The considered detector types and their most important parameters.}
\label{tab:detectors}
\end{table}
The considered beams and detectors allow now different interesting
combinations which are listed in table~\ref{tab:scenarios}. JHF-SK
is the planned combination of the existing SuperKamiokande detector
and the JHF beam, while JHF-HK is the combination of an upgraded 
JHF beam with the proposed HyperKamiokande detector. With typical
parameters, JHF-HK is altogether about $95$ times more integrated 
luminosity than JHF-SK, and we assume that it operates partly with 
the anti-neutrino beam. Water Cherenkov detectors are ideal for the
JHF beam, since charged current quasi elastic scattering is dominating.
NuMI is the proposed combination of the NuMI off-axis beam with a 
low-Z calorimeter, which is better here, since the energy is higher
and there is already a considerable contribution of inelastic 
charged current interactions. 
NuFact-I is an initial neutrino factory, while NuFact-II is a fully 
developed machine, with $42$ times the luminosity of NuFact-I 
\cite{Itow:2001ee,Para:2001cu,Blondel:2000gj}. Deep inelastic scattering
dominates for these even higher energies and magnetized iron 
detectors are therefore considered in combination with neutrino factories.

\begin{table}
\bc
\begin{tabular}{|l|c|c|c|c|}
\hline
acronym & detector & baseline & matter density & $L/E_{\mathrm{peak}}$\\
\hline\hline
{\bf JHF-SK~} & water Cherenkov & $295\,\mathrm{km}$ & 
$2.8\,\mathrm{g}\,\mathrm{cm}^{-3}$ & $378\,\mathrm{km}\,\mathrm{GeV}^{-1}$ 
\\ 
\hline
{\bf NuMI~} & low-Z & $735\,\mathrm{km}$ & 
$2.8\,\mathrm{g}\,\mathrm{cm}^{-3}$ & $337\,\mathrm{km}\,\mathrm{GeV}^{-1}$ 
\\ 
\hline
{\bf NuFact-I~} & $10$~kt magnetized iron & $3\,000\,\mathrm{km}$ & 
$3.5\,\mathrm{g}\,\mathrm{cm}^{-3}$ & $90\,\mathrm{km}\,\mathrm{GeV}^{-1}$ 
\\ 
\hline
{\bf JHF-HK~} & water Cherenkov & $735\,\mathrm{km}$ & 
$2.8\,\mathrm{g}\,\mathrm{cm}^{-3}$ & $295\,\mathrm{km}\,\mathrm{GeV}^{-1}$ 
\\ 
\hline
{\bf NuFact-II~} & $40$~kt magnetized iron & $3\,000\,\mathrm{km}$ & 
$3.5\,\mathrm{g}\,\mathrm{cm}^{-3}$ & $90\,\mathrm{km}\,\mathrm{GeV}^{-1}$ 
\\
\hline
\end{tabular}
\ec
\caption{The considered combinations of beams and detectors and their acronyms.}
\label{tab:scenarios}
\end{table}

\begin{table}
\bc
\begin{tabular}{|l|c|c|c|c|c|}
\hline
&{\bf ~JHF-SK~} & {\bf ~NuMI~} & {\bf ~JHF-HK~} & {\bf ~NuFact-I~} & {\bf ~NuFact-II~}\\
\hline\hline
signal & $139.0$ & $387.5$ & $13\,180.0$ & $1\,522.8$ & $64\,932.6$\\
\hline
background~  & $23.3$ & $53.3$ & $2\,204.6$ & $4.2$ & $180.3$\\
\hline
S/N & $6$ & $6$ & $6$ & $360$ & $360$\\
\hline
\end{tabular}
\ec
\caption{The expected signal and background event rates for the 
appearance channels for the considered scenarios.}
\label{tab:ratesap}
\end{table}


\section{The Qualitative Picture}
\label{sec:Qpicture}

\begin{figure}[tb!]
\bc
\includegraphics[width=13.5cm,angle=0]{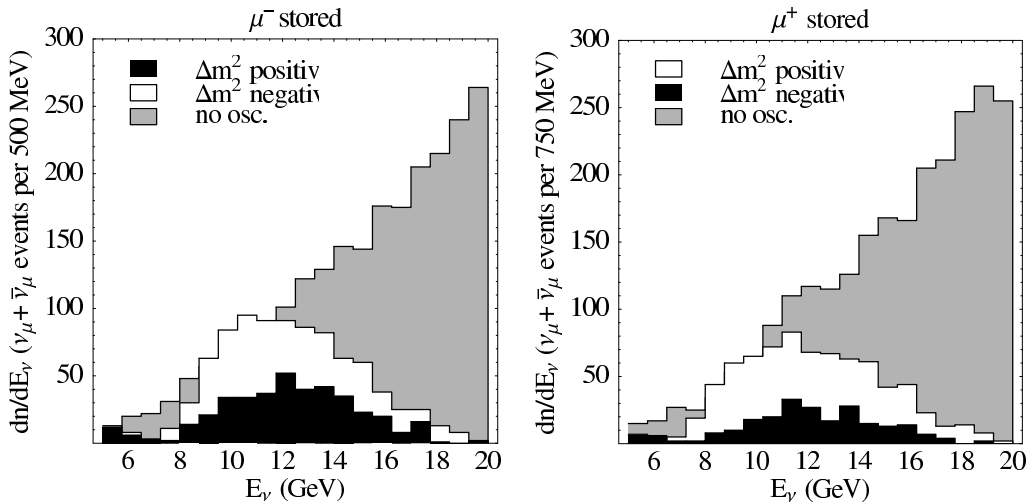}  
\ec
\caption{$\nu_\mu$ disappearance rates for a typical neutrino factory setup.
The grey area marks the expected distribution in the absence of oscillations.
The black and white histograms show the expected distributions with
oscillations for both $\dm{31}>0$ and $\dm{31}<0$. Here a relatively large 
$\sin^2 2\theta_{13}= 0.01$ was chosen, where matter effects allow to 
extract $sign(\dm{31})$ even from the disappearance channel, which is not 
possible for smaller $\sin^2 2\theta_{13}$ \cite{Freund:2000ti}.}
\label{fig:DISex}
\end{figure}

A number of studies have analyzed various aspects of individual or
some combined variants of the above scenarios 
\cite{Albright:2000xi,Cervera:2000kp,Freund:2001ui,Barger:2001yr,
Minakata:2001qm,DeRujula:1998hd,Barger:1999fs,Freund:1999gy,
Mocioiu:2000st,Barger:2000yn,Freund:2000ti,Bueno:2000fg,
Donini:2000ky,Barger:2000nf,Campanelli:wi,Freund:vt,Mocioiu:2001jy,
Yasuda:2001ip,Barger:2001yx,Bueno:2001jd,Yasuda:2002jk,
Barenboim:2002zx,Alsharoa:2002wu,Huber:2002mx} and the results can
all be understood by the same qualitative picture.
The proposed setups lead in general to remarkably large event rates
in the disappearance channel. This leads to many events per energy 
bin, and the spectral information allows very precise fits of 
the leading oscillation parameters $\dm{31}$ and $\theta_{23}$.
A typical example is shown for a neutrino factory in fig.~\ref{fig:DISex}
for both polarities \cite{Freund:2000ti}. 
The situation is somewhat different in the appearance channels, 
where the event rates are small. The results depend thus for 
the appearance channels dominantly on total rates, with 
some spectral information. Note, however, that available spectral 
information is very important, since it allows to distinguish 
solutions which are degenerate on the basis of total event rates. 
\begin{figure}[htb!]
\bc
\includegraphics[width=5.2cm,angle=-90]{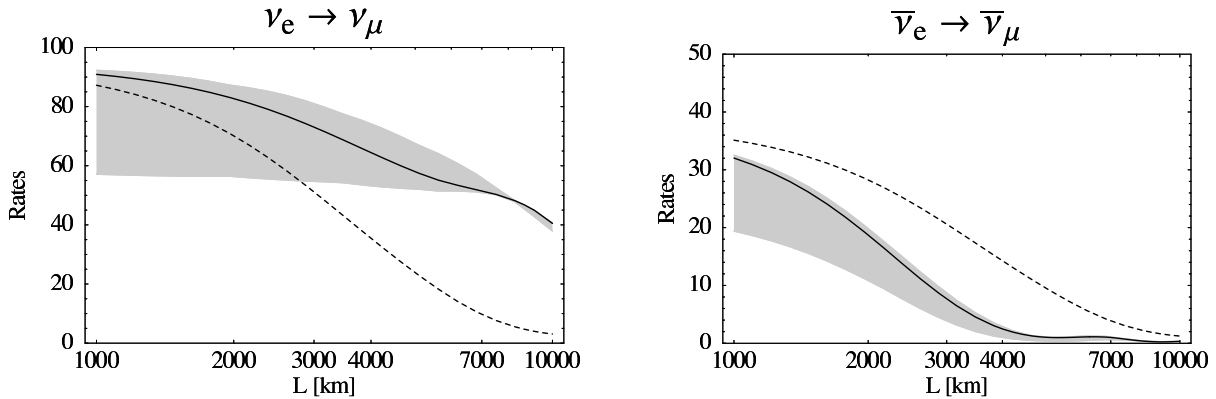}  
\ec
\caption{The total event rates for the two polarities of a 
neutrino factory with $E_\mu=50~$GeV and $\sin^2 2\theta_{13}=0.01$ 
and the LMA solution. The solid lines are for $\delta=0$ with the
matter corrections included. The dashed lines show for comparison
$\delta=0$ in vacuum. The grey band shows the range where the solid
line moves when the CP phase is allowed to take all possible values
with matter included.}
\label{fig:APex}
\end{figure}
Fig.~\ref{fig:APex} shows how the total event rates depend for a 
neutrino factory with typical parameters on the presence of matter 
and on the CP phase. It can clearly be seen in fig.~\ref{fig:APex}
that matter effects grow with distance and become dominating
at large baselines of $L\gta 3000~\mathrm{km}$. Contrary the 
CP phase $\delta$ affects the rates at shorter distances, while
at medium distances comparable matter effects and CP violating effects 
are present. The simplest strategy to separate matter effects 
and CP violation is thus to have one short baseline for CP violating
effects and another large baseline for matter effects
\cite{Cervera:2000kp,Barger:2002rr}. Alternatively one might use 
a single medium baseline, where the effects can be separated if 
the event rates are large enough such that spectral information 
can be extracted \cite{Freund:2001ui}. Another alternative 
is, as discussed before, to use one baseline and further oscillation 
channels \cite{Donini:2002rm}.

It is interesting to understand the interplay of matter effects and CP
violating effects from our analytic formulae. First we observe from 
eqs.~(\ref{eq:Pdis}) and (\ref{eq:Pap}) as well as fig.~\ref{fig:APex} 
that matter effects grow with distance and CP violating effects are thus 
not affected by matter effects at shortest distances. Inspecting 
carefully eq.~(\ref{eq:Pap}) we found in section~\ref{sec:CD} for larger 
distances the existence of a magic baseline. The point was that 
$\hat A = 2VE/\dm{31}$ and $\Delta=\dm{31}L/E$, such that 
$\hat A\Delta = 2VL$ depends only on the matter potential $V$
and the baseline $L$. All but the first term in eq.~(\ref{eq:Pap})
vanish therefore for $L_{magic}$ as given in eq.~(\ref{eq:Lmagic}), 
since they contain factors of $\sin(\hat A\Delta)=\sin(2VL)$. 
All terms which contain the CP phase $\delta$ and the mass splitting 
hierarchy parameter $\alpha$ vanish thus as a consequence of matter 
effects at $L_{magic}\simeq 8100~\mathrm{km}$, where a typical density 
in the Earth crust has been used. Matter effects and CP violating effects 
can thus in principle be completely separated by performing  
measurements at small baseline and at the magic baseline \cite{Hubertalk}. 

\section{The Analysis}
\label{sec:analysis}

The sensitivity and the precision of the measured quantities are 
defined as explained in section~\ref{sec:evrates} via the ability 
to re-extract the physics parameters from 
previously generated event rate distributions. In order to 
determine the sensitivity and the precision all input parameters 
are scanned while systematical, statistical and background errors 
are included. The best possible results are obtained by optimally 
using the information contained in the rates and in the energy 
spectrum of all available channels in a combined analysis. At the 
same time the parameter correlations and degeneracies mentioned 
above must be included in addition to systematical errors like 
normalization and calibration. However, precise relative information 
contained in the spectrum allows very precise measurements even 
for an included overall normalization error of $5\%$ and an energy 
calibration error of $5\%$. The inclusion of backgrounds limits the 
sensitivity. However, information in the energy spectrum helps to 
reduce the impact of the background, which has typically a 
different energy dependence. 

There are many events per bin in the disappearance channels, which
leads via eq.~(\ref{eq:Pdis}) to a very precise determination of 
the leading oscillation parameters $\dm{31}$ and $\sin^2 2\theta_{23}$.
The combination of the available appearance channels allows to 
determine or restrict $\theta_{13}$, and via the matter effects 
$sign(\dm{31})$. This is always possible, even when $\alpha$ is 
tiny or negligible for the disfavoured SMA, LOW and VAC solutions 
of the solar neutrino problem, since the first term in eq.~(\ref{eq:Pap}) 
does not depend on $\alpha$.
For the LMA solution (\ie\ $\alpha\simeq 10^{-2}$) it is possible
to determine $\theta_{13}$, $sign(\dm{31})$, and it is in principle 
even possible to extract from a combined fit of the appearance and 
disappearance channels the solar parameters $\theta_{12}$ and 
$\dm{21}$ without using external input. The precision which 
can be obtained for $\theta_{12}$ and $\dm{21}$ can, however, not 
compete with the expected measurements of KamLand. We assume 
therefore in our analysis that KamLand measures the solar parameters 
at the current best fit of the LMA solution with typical errors and 
take this as external input for our analysis. This allows then for 
the favoured LMA solution to extract information on the CP phase 
$\delta$, as expected from the second and third term of eq.~(\ref{eq:Pap}). 

\begin{figure}[htb!]
\begin{center}
\includegraphics[width=13.5cm]{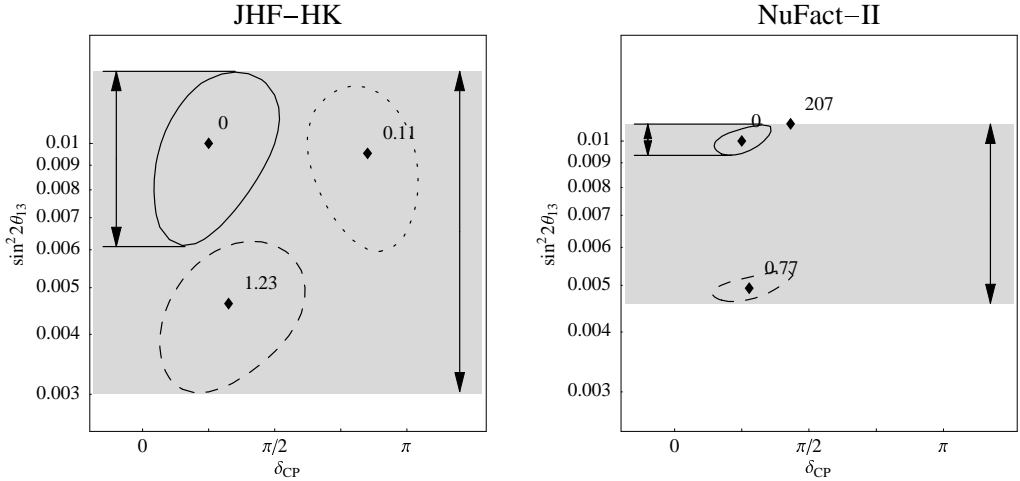} 
\end{center}
\caption{Examples of degenerate solutions \cite{Huber:2002mx}. The left 
plot shows a case where the degenerate solutions are very close. The 
combination of these adjacent solutions leads consequently to an enlarged 
error due to degeneracies. The right plot shows a case where two degenerate 
solutions are well separated. In this case two solutions and their 
respective errors should be quoted.}
\label{fig:degcorr}
\end{figure}

It should be kept in mind that eqs.~(\ref{eq:Pdis}) and 
(\ref{eq:Pap}) exhibit parameter correlations and degeneracies as 
discussed above, which must be taken into account in the analysis
\cite{Cervera:2000kp,Barger:2001yr,Burguet-Castell:2001ez,
Minakata:2001qm,Freund:2001ui,Huber:2002mx}.
The analytic formulae eqs.~(\ref{eq:Pdis}) and (\ref{eq:Pap}) are
here extremely useful, since they allow to understand qualitatively
the highly non-linear behaviour and the complex topology of the 
parameter manifolds. The parameter dependence in the probabilities 
leads, as discussed above, to three degeneracies in the following 
parameter planes:
\begin{itemize}
\item $\delta$-$\theta_{13}$
\item $\delta$-$\mathrm{sign}(\dm{31})$
\item $\theta_{23}$-- ($\pi/2-\theta_{23}$)
\end{itemize}
These three different degeneracies can lead to equivalent solutions 
which can be rather close in parameter space. In this case they effectively
enlarge the allowed range of the combined solutions as shown in an 
example for the $\delta-\theta_{13}$ plane in the left plot of 
fig.~\ref{fig:degcorr}. In other cases the degenerate solutions 
are well separated, as it is shown, for example, in the right plot of 
fig.~\ref{fig:degcorr}. In such cases it is more sensible to quote two 
values and their respective errors.
In a detailed single baseline analysis one can see, as discussed 
analytically in section~\ref{sec:CD}, that the $\delta$-$\theta_{13}$ 
degeneracy becomes typically a parameter correlation as long as only a 
neutrino beam is used, while degenerate islands show up when both 
polarities are combined. Similarly one can see that with one baseline 
the $\delta$-$\mathrm{sign}(\dm{31})$ degeneracy can never be removed.
The $\theta_{23}$--($\pi/2-\theta_{23}$) degeneracy can also not be removed
as long as the analysis is dominated by the disappearance rates only.
This degeneracy can be lifted for sufficiently high statistics in the 
appearance channel, where the degeneracy is in principle broken, as
seen in the analytic discussion in section~\ref{sec:CD}.

\begin{figure}[htb!]
\bc
\includegraphics[width=6cm,angle=-90]{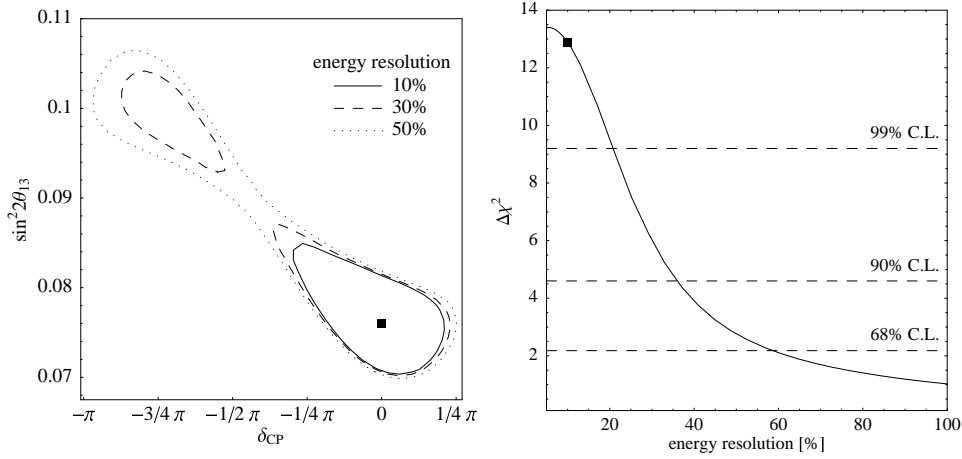}
\ec
\caption{The correlation or degeneracy of $\delta$ with $\theta_{13}$
as a function of the detector energy resolution for a neutrino
factory with $E=50\,\mathrm{GeV}$, $L=3\thinspace 000\,\mathrm{km}$, 
$\dm{31}=3.5\cdot 10^{-3}\,\mathrm{eV}^2$, $\dm{21}=10^{-4}\,\mathrm{eV}^2$ 
and $\theta_{23}=\theta_{12}=\pi/4$ \cite{Freund:2001ui}.
For poor resolution there is a correlation between $\delta$ with 
$\theta_{13}$ which transforms into two degenerate solutions for 
medium resolution, and the degeneracy is finally lifted for energy 
resolution better than 25\%.}
\label{fig:degDchi}
\end{figure}

The separation of matter and CP violating effects is, as discussed
in section~\ref{sec:CD}, not possible at the level of total
event rates for one value of $L/E$, which translates for fixed $E$ 
into one baseline. However, as mentioned before, there exist ways to 
break this degeneracy. One strategy was to use a short baseline
for CP violating effects and a long baseline for matter effects.
Another strategy was to use one single medium baseline and the 
information contained in the beam spectrum. The point is that 
different parameter sets were degenerate at the level of total 
event rates for one $L/E$, while the event rate distributions differ 
significantly. The degeneracy may thus be lifted for one medium 
baseline in combination with good resolution and sufficient 
statistics. This is shown in fig.~\ref{fig:degDchi} for the 
correlation of $\delta$ with $\theta_{13}$ \cite{Freund:2001ui}.
This example shows nicely how degeneracies and correlations 
can change in a real experiment at the level of event rates
when the spectral information is used. 

\section{Results}
\label{sec:results}

A realistic analysis of future LBL experiments requires a number 
of different aspects to be taken into account. It should be clear 
from the discussion above that it is not sufficient to quote limits 
which are based on oscillation probabilities or merely on the 
statistics of a single channel without backgrounds or systematics. 
Depending on the position in the space of physics parameters the 
degeneracies or correlations, the backgrounds, the systematics or 
statistics may be the limiting factor. A reliable comparative 
study of the discussed LBL setups requires therefore a detailed 
analysis of the six dimensional parameter space, which 
includes all these effects on the same footing \cite{Huber:2002mx}.

\begin{figure}[htb!]
\hspace*{30mm}{\tiny\bf II$\leftarrow$I}\hspace*{22mm}{\tiny\bf HK$\leftarrow$SK}
\hspace*{40mm}{\tiny\bf II$\leftarrow$I}\hspace*{17mm}{\tiny\bf HK$\leftarrow$SK}
\hfill\\[-12mm]
\bc
\includegraphics[width=7.0cm,angle=-90]{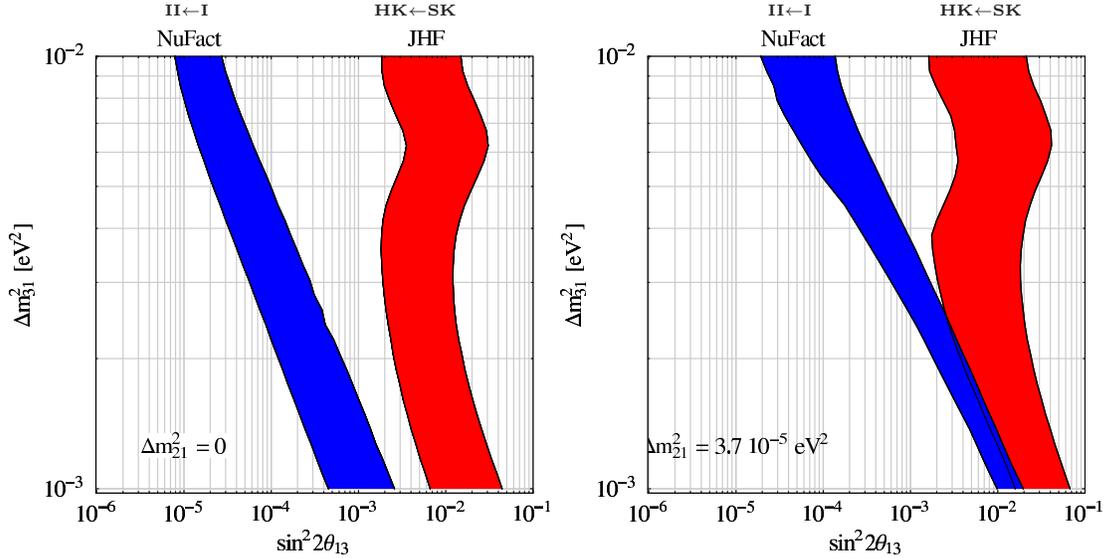} 
\ec
\caption{The sensitivity to $\sin^22\theta_{13}$ for the considered JHF and
NuFact setups as a function of $\dm{31}$ for $\dm{21}=0$ (left plot) and 
$\dm{21}=3.7~10^{-5}~\mathrm{eV}^2$ (right plot). The right edge of the
NuFact and JHF bands corresponds to the less advanced options, NuFact~I 
and JHF-SK, while NuFact~II and JHF-HK are equivalent to the left side of 
the bands \cite{Huber:2002mx}.}
\label{fig:th13sens}
\end{figure}

There is excellent precision for the leading oscillation parameters
$\dm{31}$ and $\sin^2 2\theta_{23}$, which will not be further discussed 
here. The more interesting sensitivity to the sub-leading parameter 
$\sin^2 2\theta_{13}$ is shown in fig.~\ref{fig:th13sens},
where we can see that the $\sin^22\theta_{13}$ sensitivity limit depends 
considerably on the value of $\dm{31}$. From the comparison of the cases 
$\dm{21}=0$ (left plot) and  $\dm{21}=3.7~10^{-5}~\mathrm{eV}^2$ (right 
plot) we find moreover a significant $\dm{21}$ dependence of the 
sensitivity limits\footnote{This translates into a corresponding
$\dm{31}$ and $\dm{21}$ dependence of the obtainable precision in 
case $\sin^22\theta_{13}$ is large enough to be measured. The 
corresponding plots can be found in \cite{Huber:2002mx}.}. 
The $\sin^22\theta_{13}$ sensitivity limit can thus change considerably
depending on what will be found for $\dm{31}$ and $\dm{21}$ and 
these dependencies are strongest for short baselines.

\begin{figure}[tb]
\bc
\includegraphics[width=12cm]{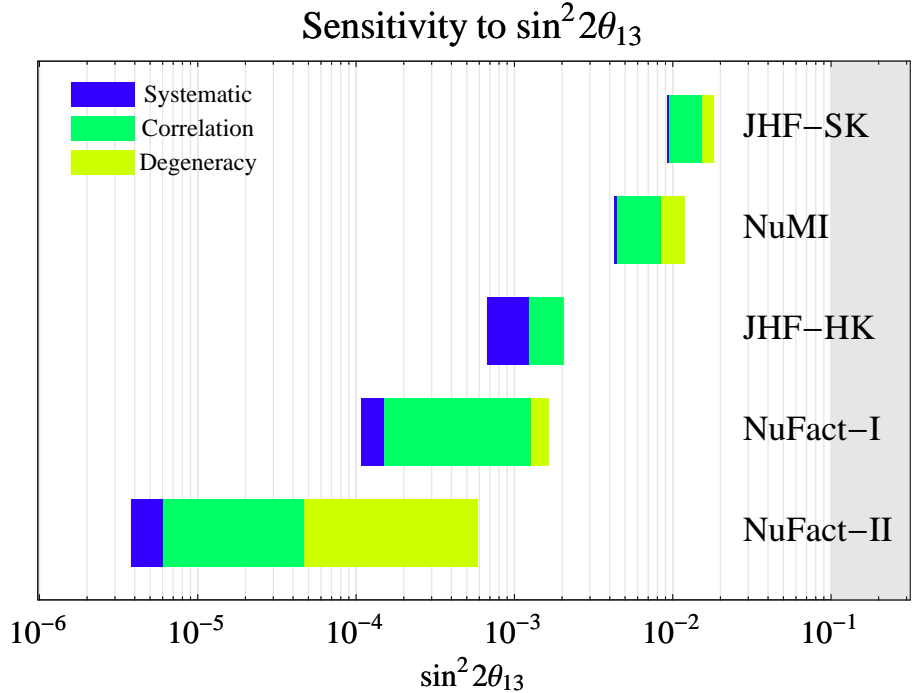} 
\ec
\caption{The $\sin^2 2 \theta_{13}$ sensitivity for all setups 
defined in section~\ref{sec:LBLpot} at the 90\%  confidence level for 
$\dm{31}= 3\cdot 10^{-3} \, \mathrm{eV}^2$ and $\sin^2 2 \theta_{23}=0.8$.
The plot shows the deterioration of the sensitivity limits as the
different error sources are successively switched on. The left
edge of the bars is the sensitivity statistical limit. This 
limit gets reduced as systematical, correlational and degeneracy
errors are switched on. The right edge is the final sensitivity 
limit \cite{Huber:2002mx}.}
\label{fig:th13exec}
\end{figure}

Assuming that the leading parameters are measured to be 
$\dm{31}= 3\cdot 10^{-3}~\mathrm{eV}^2$, $\sin^2 2 \theta_{23}=0.8$
and that KamLand measures the solar parameters at the current
best fit point of the LMA region, \ie\ $\dm{21}= 6\cdot 10^{-5}~\mathrm{eV}^2$ 
and $\sin 2\theta_{12} = 0.91$, we can translate this into a comparison of the 
$\sin^2 2\theta_{13}$ sensitivity limit for the different setups. The 
result is shown in fig.~\ref{fig:th13exec}. The individual contributions
of different sources of uncertainties are shown for every experiment
and the left edge of every band in fig.~\ref{fig:th13exec} corresponds 
to the sensitivity limit which would be obtained purely on statistical 
grounds. This limit is successively reduced by adding the systematical 
uncertainties of each experiment, the correlational errors and finally 
the degeneracy errors. The right edge of each band constitutes the 
final error for the experiment under consideration.
It is interesting to see how the errors of the different setups are
composed. There are different sensitivity reductions due to systematical
errors, correlations and degeneracies.
The largest sensitivity loss due to correlations and degeneracies 
occurs for NuFact-II, which is mostly a consequence of the 
uncertainty of $\dm{21}$, which translates into an $\alpha$ uncertainty,
and which dominates the appearance probability eq.~(\ref{eq:Pap}) for 
small $\sin^22\theta_{13}$. Note that it is in principle
possible to combine different experiments. If done correctly, this 
allows to eliminate part or all of the correlational and degeneracy 
errors \cite{Burguet-Castell:2002qx}.

\begin{figure}[tb]
\bc
\includegraphics[width=12cm]{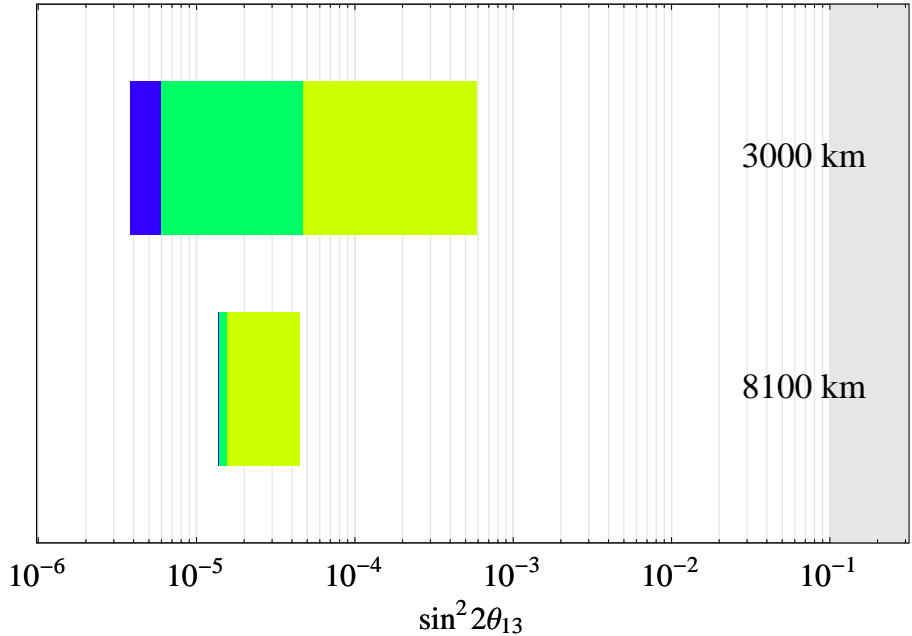} 
\ec
\caption{Comparison of the NuFact-II setup for a baseline of 3000~km
and the magic baseline of 8100~km. It can be seen that the statistical
sensitivity (left edge of the bars) is reduced due to smaller event 
rates, but the total sensitivity is increased since the correlational 
and degeneracy errors disappear almost completely at the magic baseline.}
\label{fig:th13magic}
\end{figure}

It is interesting to recall the existence of the magic baseline
discussed in eq.~(\ref{eq:Lmagic}), where the $\alpha$ and $\delta$ 
dependence drops out completely due to matter effects. Correlational 
and degeneracy errors are then drastically reduced and a measurement 
of $\sin^22\theta_{13}$ becomes more precise, even though the event 
rates are smaller at this larger baseline. This improvement in 
sensitivity is shown for NuFact-II in fig.~\ref{fig:th13magic}, 
where a baseline of 3000~km is compared with the magic baseline of 
$8100~\mathrm{km}$.

\begin{figure}[tb]
\bc
\includegraphics[width=9.0cm,angle=-90]{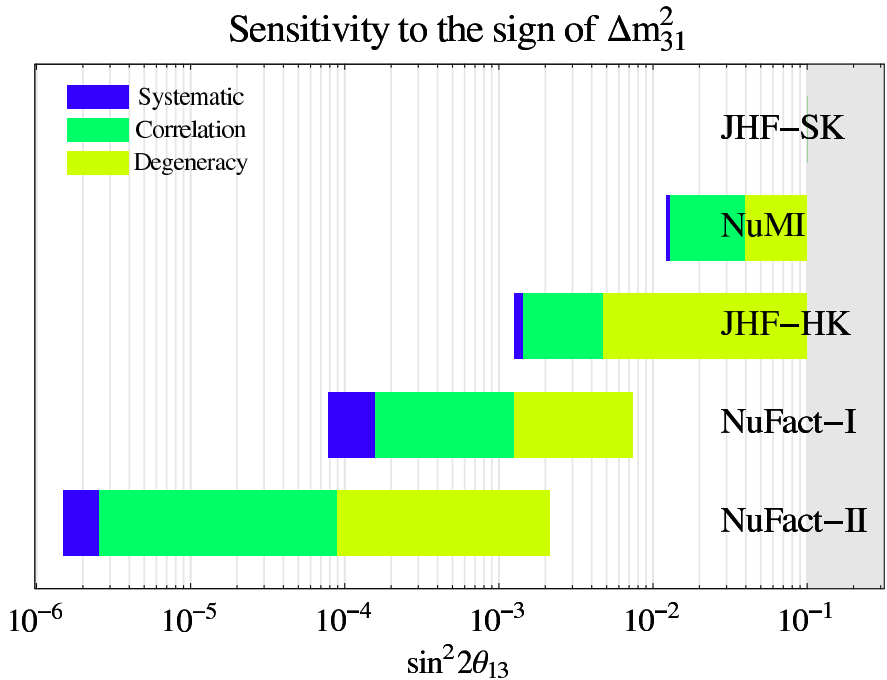} 
\ec
\caption{The $\sin^2 2\theta_{13}$ sensitivity region to 
$sign(\dm{31})$ for the setups defined in section~\ref{sec:LBLpot}.
The left edge of the bars are the statistical sensitivity limits
which are successively reduced by systematical, correlational and
degeneracy errors. The right edge of the bars is the final limit.}
\label{fig:signex}
\end{figure}

Another challenge of future LBL experiments is to measure $sign(\dm{31})$ 
via matter effects and the sensitivity which can be obtained for the setups 
under discussion is shown in fig.~\ref{fig:signex}.
Taking all correlational and degeneracy errors into account we 
can see that it is very hard to determine $sign(\dm{31})$ with 
the considered superbeam setups. The main problem is the degeneracy 
with $\delta$, which allows always the reversed $sign(\dm{31})$ 
for another CP phase. Note, however, that the situation can in
principle be improved if different superbeam experiments were 
combined such that this degeneracy error could be removed. 
Neutrino factories perform considerably better on $sign(\dm{31})$,
particularly for larger baselines. Combination
strategies would again lead to further improvements.

Coherent forward scattering of neutrinos and the corresponding 
MSW matter effects are so far experimentally untested. It is 
therefore very important to realize that matter effects will not 
only be useful to extract $sign(\dm{31})$, but that they allow 
also detailed tests of coherent forward scattering of neutrinos. 
This has been studied in detail in \cite{Freund:2001ui,Freund:1999gy,
Freund:2000ti,Freund:vt}.

The Holy Grail of LBL experiments is the measurement of leptonic
CP violation.
\begin{figure}[tb]
\bc
\includegraphics[width=8.4cm,angle=-90]{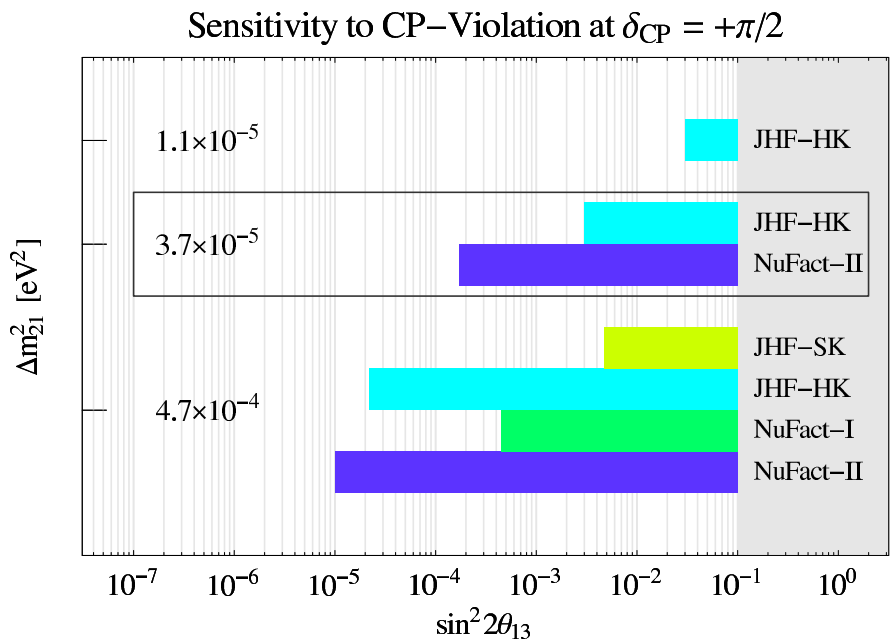} 
\ec
\caption{The $\sin^2 2 \theta_{13}$ sensitivity range for CP violation 
of the considered setups at 90\% confidence level and for different
$\Delta m_{21}^2$ values. The upper row corresponds to the lower bound 
of $\Delta m_{21}^2=1.1\times 10^{-5}~\mathrm{eV}^2$, the bottom row
to the upper bound $\Delta m_{21}^2=4.7\times 10^{-4}~\mathrm{eV}^2$,
and the middle row to the best LMA fit, 
$\Delta m_{21}^2=3.7\times 10^{-5}~\mathrm{eV}^2$. 
Cases which do not have CP sensitivity are omitted from this plot.
The chosen parameters are $\delta=+\pi/2$, 
$\dm{31}= 3\cdot 10^{-3}~\mathrm{eV}^2$, $\sin^2 2 \theta_{23}=0.8$,
and a solar mixing angle corresponding to the current best fit
in the LMA regime \cite{Huber:2002mx}.}
\label{fig:CPexec}
\end{figure}
The $\sin^2 2\theta_{13}$ sensitivity range for measurable CP violation 
is shown in fig.~\ref{fig:CPexec} for $\delta=\pi/2$ for the different
setups and for different values of $\dm{21}$. It can be seen that 
measurements of CP violation are in principle feasible both with high 
luminosity superbeams as well as advanced neutrino factories. However, 
the sensitivity depends in a crucial way on $\dm{21}$. For a low value 
$\dm{21}=1.1~10^{-5}~\mathrm{eV}^2$, the sensitivity is almost completely 
lost, while the situation would be very promising for the largest considered 
value $\dm{21}=4.7~10^{-4}~\mathrm{eV}^2$. For a measurement of leptonic 
CP violation it would therefore be extremely exciting and promising if 
KamLand would find $\dm{21}$ on the high side of the LMA solution (the 
so-called HLMA case). 
The sensitivities shown in fig.~\ref{fig:CPexec} depend on the choice
for $\delta$. The value which was used here was $\delta=\pi/2$ and
and the sensitivities become become worse for small CP phases close to 
zero or $\pi$.

\section{Conclusions}
\label{sec:Conclusio}

We have discussed the potential of certain future long baseline 
neutrino oscillation experiments, where it will be possible to 
perform precision neutrino physics. The basic fact which makes
this possible is that the atmospheric mass splitting
$\dm{31}\simeq\Delta{m^2_\mathrm{atm}}$ leads for typical neutrino 
energies $E_\nu\simeq 1-100$~GeV to oscillation baselines in 
the range $100$~km to $10000$~km. Beam sources have moreover 
the advantage, that unlike the sun or the atmosphere, they 
can be controlled very precisely, such that unknowns of the 
neutrino source do not limit the precision. Equally precise
detectors and an adequately precise oscillation framework (including 
three neutrinos and matter effects) must be used in order to 
exploit this precision. There exist other interesting sources
for long baseline oscillation experiments, like reactors or 
$\beta$-beams, but we restricted the discussion here to superbeams 
and neutrino factories. We presented the issues which enter into 
realistic assessments of the potential of such experiments. 
The discussed experiments turned out to be very promising and 
they lead to very precise measurements of the leading oscillation
parameters $\dm{31}$ and $\sin^2 2\theta_{23}$. We discussed
in detail how the different setups lead to very interesting 
measurements or limits $\theta_{13}$ and $\delta$. It will also 
be possible to perform impressive tests of Earth matter effects, 
allowing to extract $sign(\dm{31})$. The discussed setups have
an increasing potential and increasing technological challenges,
but it seems possible to built them in stages. The shown results
are valid for each individual setup and future results should
of course be included in the analysis. This would be especially
important if more LBL experiments were built and depending on
previous results there exist different optimization strategies.
In the short run the expected results from KamLand are extremely
important and will have considerable impact. First it will 
become clear if $\dm{21}$ lies in the LMA regime, which is 
very important since realistically, CP violation can only be 
measured then. Within the LMA solution it is also very important
if $\dm{21}$ lies close to the current best fit, on the high or 
on the low side. A value of $\dm{21}$ on the high side (HLMA) would
be ideal, since it would guarantee an extremely promising LBL
program with a chance to see leptonic CP violation already with 
the JHF beam in the next decade.

Finally we would like to stress that the physics program of LBL 
experiments has a unique impact on physics. It would lead to 
very precise neutrino mass splittings and very precise leptonic 
mixings. These measurements yield directly the physics parameters 
of interest, which are (unlike quarks) not masked by any hadronic 
uncertainties. It would thus be possible to get very valuable 
lepton flavour information, which could be directly compared 
with models of masses and mixings as well as renormalization 
group effects. Such precise leptonic flavour information might 
also proof more valuable than in the quark sector, since neutrino 
masses receive in general contributions both from Dirac and 
Majorana mass terms and more might be learned in this way.
Leptonic CP violation is also an extremely interesting issue, 
since it is related to leptogenesis, which is currently the most 
plausible mechanism to explain the baryon asymmetry of the universe.
The experiments which were presented here allow also a number of other 
studies which have not been discussed here. Some examples are 
limits on non-standard interactions, FCNC, more than three 
neutrinos, CPT violation. The results presented here should,
however, make clear that oscillation physics with long 
baseline experiments alone is already very interesting, 
powerful and important. The realization of the discussed setups 
is not easy, but it appears possible in stages, guaranteeing a 
very promising future of neutrino physics.

\vs

\noindent
{\bf Acknowledgments:} I would like to thank 
M.~Freund, P.~Huber and W.~Winter for the collaboration in 
the studies on which this article is based upon.
This work was supported by the ``Sonderforschungsbereich 375 
f\"ur Astro-Teilchenphysik'' der Deutschen Forschungsgemeinschaft.


%
%


\end{document}